\documentclass[useAMS,usenatbib]{mn2e}
\usepackage[dvips]{graphicx}

\title[The Incidence of Growing Black Holes]
  {Towards a Complete Census of AGNs in Nearby Galaxies: The Incidence of Growing Black Holes}
\author[A. Goulding et al.]
  {A.D.~Goulding$^1$\thanks{E-mail:- andrew.goulding@durham.ac.uk}
    D.M.~Alexander$^1$ 
    B.D.~Lehmer$^{\rm 1,2,3}$
    J.R.~Mullaney$^1$\\
  $^1$Department of Physics, Durham University, South Road, Durham DH1 3LE, UK\\
  $^2$The Johns Hopkins University, Homewood Campus, Baltimore, MD 21218, USA\\
  $^3$NASA Goddard Space Flight Centre, Code 662, Greenbelt, MD 20771, USA}

\date{Released 2010 Xxxxx XX}

\pagerange{\pageref{firstpage}--\pageref{lastpage}} \pubyear{2008}

\def\LaTeX{L\kern-.36em\raise.3ex\hbox{a}\kern-.15em
    T\kern-.1667em\lower.7ex\hbox{E}\kern-.125emX}

\def\spitzer{{\it Spitzer}}

\def\cm{{\rm\thinspace cm}}

\def\erg{{\rm\thinspace erg}}

\def\km{{\rm\thinspace km}}

\def\Lsun{\hbox{$\rm\thinspace L_{\odot}$}}

\def\Msun{\hbox{$\rm\thinspace M_{\odot}$}}

\def\s{{\rm\thinspace s}}

\def\ergps{\hbox{$\erg\s^{-1}\,$}}

\def\kmps{\hbox{$\km\s^{-1}\,$}}

\def\pcmsq{\hbox{$\cm^{-2}\,$}}
\def\Mbh{\hbox{${\rm M}_{\rm BH}$}}

\def\Lir{\hbox{${\rm L}_{\rm IR}$}}

\def\Lbul{\hbox{${\rm L}_{\rm bul}$}}

\def\um{\hbox{$\ \umu {\rm m}$}}
\def\nev{[NeV] $\lambda 14.32 \um$ }

\begin{document}

\label{firstpage}

\maketitle

\begin{abstract}
We investigate the local supermassive black hole (SMBH) density
function and relative mass accretion rates of all active galactic
nuclei (AGNs) identified in a volume-limited sample of infrared (IR)
bright galaxies ($\Lir > 3 \times 10^9 \Lsun$) to $D<15$~Mpc (Goulding
\& Alexander 2009). A database of accurate SMBH mass ($\Mbh$)
estimates is compiled from literature sources using physically
motivated AGN modeling techniques (reverberation mapping, maser
mapping and gas kinematics) and well-established indirect $\Mbh$
estimation methods (the M-$\sigma_*$ and $\Mbh$--$L_{\rm K,bul}$
relations). For the three sources without previously published $\Mbh$
estimates, we use 2MASS $K$-band imaging and {\sc galfit} to constrain
the bulge luminosities, and hence SMBH masses. In general, we find the
AGNs in the sample host SMBHs which are spread over a wide mass range
($\Mbh \approx (0.1$--$30) \times 10^7 \Msun$), but with the majority
in the poorly studied $\Mbh \approx 10^6$--$ 10^7 \Msun$ region. Using
sensitive hard X-ray (2--10 keV) and mid-IR constraints we calculate
the bolometric luminosities of the AGNs ($L_{\rm Bol,AGN}$) and use
them to estimate relative mass accretion rates. We use these data to
calculate the volume-average SMBH growth rate of galaxies in the local
Universe and find that the AGNs hosting SMBHs in the mass range $\Mbh
\approx 10^6$--$10^7 \Msun$ are dominated by optically unidentified
AGNs. These relatively small SMBHs are acquiring a significant
proportion of their mass in the present-day, and are amongst the most
rapidly growing in the local Universe (SMBH mass doubling times of
$\approx$~6~Gyrs). Additionally, we find tentative evidence for an
increasing volume-weighted AGN fraction with decreasing SMBH mass in
the $\Mbh \approx 10^6$--$10^8 \Msun$ range. Overall, we conclude that
significant mass accretion onto small SMBHs may be missed in even the
most sensitive optical surveys due to absent or weak optical AGN
signatures.

 \end{abstract}

\begin{keywords}
galaxies: active -- galaxies: evolution -- galaxies: nuclei --
infrared: galaxies
\end{keywords}

\section{Introduction}

It is now well established that all massive galaxies ($M_* \approx
10^{10}$--$10^{12} \Msun$) in the local Universe harbour central
super-massive black holes (SMBHs), with masses proportional to those
of their stellar spheroids (hereafter, bulge;
e.g.,\ \citealt{kormendy95}; \citealt{magorrian98}). Comparisons
between the SMBH mass density in the local Universe and the total
energy produced by active galactic nuclei (AGNs) across cosmic time
have shown that these SMBHs were primarily grown through
mass-accretion events (e.g.,\ \citealt{soltan82}; \citealt{rees84};
\citealt{marconi04}). The space density of high-luminosity AGNs
appears to have peaked at higher redshifts than lower-luminosity AGNs,
suggesting that the most massive SMBHs ($\Mbh \approx 10^8$--$10^9
\Msun$) grew first, a result commonly referred to as `AGN cosmic
downsizing' (e.g.,\ \citealt{cowie03}; \citealt{ueda03};
\citealt{mclure04}; \citealt{hasinger05};
\citealt{alonso08}). Extrapolation of these results imply that the
most rapidly growing SMBHs in the nearby Universe should be of
comparatively low mass ($\Mbh \ll 10^{8} \Msun$). To determine the
characteristic masses of these growing SMBHs requires a complete
census of AGN activity and SMBH masses in the local Universe.

Using data from the Sloan Digital Sky Survey \citep[SDSS; ][]
{sdss_tech} in conjunction with the well established SMBH--stellar
velocity dispersion relation (hereafter, M--$\sigma_*$; e.g.,
\citealt{gebhardt00}; \citealt{tremaine02}), \citeauthor{heckman04}
(2004; hereafter, H04) deduced that relatively low mass SMBHs ($\Mbh
\approx 3 \times 10^7 \Msun$) residing in moderately massive
bulge-dominated galaxies host the majority of present-day accretion
onto SMBHs. However, the space density of SMBHs derived from the
M--$\sigma_*$ relation in the optical survey of H04 was limited by the
spectral resolution of the SDSS ($ \sigma_* > 70 \kmps $) to SMBHs of
$\Mbh \goa 3 \times 10^6 \Msun$ (assuming the M--$\sigma_*$ relation
of Gebhardt et~al. 2000). Furthermore, due to attenuation of optical
emission by dust, source selection and AGN classification at optical
wavelengths will be biased against gas-rich, dust-obscured
objects. These surveys are unlikely to include galaxies hosting the
smallest bulges, and consequently the lowest mass SMBHs, and may
therefore be missing a significant proportion of SMBH growth in the
local Universe. Indeed, the nearby Scd galaxy, NGC~4945, hosting a
low-mass SMBH ($\Mbh \approx 1.4 \times 10^6 \Msun$;
\citealt{greenhill97}) only displays evidence for AGN activity in
X-ray \citep{iwasawa93} and mid-infrared (mid-IR) observations
\citep{GA09}. By contrast, the AGN in NGC~4945 (accreting at $\goa 30$
percent of the predicted Eddington limit; \citealt{itoh08}) is
completely hidden at optical wavelengths, and classified as a
starburst galaxy. Clearly, using optical data alone, the intrinsic AGN
properties of sources similar to NGC~4945 cannot be derived.

While optical emission-line diagnostics alone cannot reliably
characterise the properties of a non-negligable fraction of the AGN
population, they are readily identified at obscuration independent
wavelengths (e.g., X-ray; mid-IR). Hence, the identification of AGNs
made at X-ray and mid-IR wavelengths complements traditional
UV/optical methods to yield a more complete census of AGN
activity. Indeed, using the high resolution mid-IR spectrograph
on-board the NASA \spitzer \ Space Telescope (\spitzer-IRS),
\citeauthor{GA09} (2009; GA09) found using the first complete
volume-limited sample of all ($\approx 94$ percent) local ($D<15$ Mpc)
bolometrically luminous galaxies ($\Lir > 3 \times 10^9 \Lsun$), that
$\approx 50$ percent of local AGNs are not identified in sensitive
optical surveys. At least $30$ percent of these AGNs were previously
identified as pure optical starburst galaxies, similar to NGC~4945
(i.e., not even otherwise known to be transition-type objects as
defined by \citealt{kauff03b}). Furthermore, $\approx 30$ percent
of the optically unidentified AGNs were found to reside in late-type
spiral galaxies (Sc--Sd; e.g., similar to NGC~4945). Complimentary to
this, from a heterogenous sample of Palomar galaxies, \citet{sat07}
and \citet{sat08} have also concluded that optically unidentified AGNs
exist in some late-type spiral galaxies. With the inclusion of these
new optically unidentified AGNs, it is natural to ask, what are the
masses of local active SMBHs, what are their Eddington ratios, and
hence, how rapidly are active SMBHs growing in the local Universe?

In this paper, we investigate the growth rates and space density of
actively accreting SMBHs using the 17 AGNs identified in the
volume-limited survey of GA09. Whilst the source statistics considered
here are significantly smaller than those studies using the SDSS, this
work compliments that of H04 by including a relatively large number
(given the considered small volume) of optically unidentified AGNs
(10) which would not be reliably identified or characterised in the
SDSS survey.\footnote{Following GA09, throughout this paper we define
  an optically unidentified AGN as an object which is not
  unambiguously identified as a Seyfert galaxy at optical wavelengths
  using solely traditional optical emission-line diagnostics. This has
  the advantage that we may directly compare statistics from the
  sample considered here to those derived from large {\it N} surveys
  such as the SDSS.}  Furthermore, by including a significant
population of bolometrically luminous (but dust-obscured) late-type
spiral galaxies (Sc--Sd) we are able to extend the SMBH density
function to $\Mbh < 3 \times 10^6 \Msun$. As many of the late-type
spiral galaxies host small galactic bulges, and hence lower mass
SMBHs, particular attention is paid to obtaining accurate mass
estimates for these SMBHs. Given their proximity, many of the sources
in GA09 are well-studied and have multiple estimates of SMBH mass
($\Mbh$) from a variety of methods (i.e., reverberation mapping
techniques; mapping of water maser spots; gas kinematical estimates;
the M--$\sigma_*$ relation; correlation of $\Mbh$ with the luminosity
of the galactic bulge); below we discuss the relative accuracy of each
SMBH mass estimate technique. Furthermore, to determine the relative
mass accretion rates and hence average growth times of the SMBHs in
our sample we require the best available estimates of the AGN
bolometric luminosity ($L_{\rm Bol,AGN}$). Here we use two approaches:
1) for the AGNs with currently published data, we use high-quality
well-constrained sensitive hard X-ray (2--10 keV) luminosities to
directly measure $L_{\rm Bol,AGN}$; and 2) we accurately infer $L_{\rm
  Bol,AGN}$ using a well-constrained hard X-ray to high-ionisation
mid-IR emission line relation.

In \S2 we outline the construction and basic reduction analysis of the
AGN sample derived from GA09. In \S3 we present the SMBH mass
estimates. For a minority of objects (three out of 17 AGNs) without
published $\Mbh$ estimates we outline the use of a bulge/disc
decomposition method with 2MASS {\it K}-band images, and following
Marconi \& Hunt (2003), we use the $\Mbh$--$\Lbul$ relation to
estimate their SMBH masses. In \S4 we use hard (2--10 keV) X-ray
measurements and high-ionisation mid-IR emission to estimate the
intrinsic luminosity of the AGNs considered in our sample. Using our
well-defined estimates for SMBH mass and AGN bolometric luminosity, we
investigate the relative mass accretion rates of our sample of active
SMBHs in \S5. We use these estimates to provide new constraints on the
volume-average SMBH growth rates in the local Universe. We further
compare these results to the previous works of H04 and
\citet{gre_ho07} by producing a local AGN population density
function. Finally, in \S6 we present our conclusions.

\section{The Sample}

\subsection{Selection and Data-reduction}

The sample of local AGNs is derived from the \spitzer-IRS spectral
investigation of a volume-limited sample of IR-bright galaxies to
$D<15$ Mpc by GA09. High-resolution \spitzer-IRS spectroscopy ($R \sim
600$) was obtained for sixty-four of the sixty-eight ($\approx 94$
percent) galaxies detected in the Revised Bright Galaxy Sample (RBGS;
\citealt{RBGS}) with $\Lir \goa 3 \times 10^9 \Lsun$ within the
considered volume.\footnote{Based on the principle of reprocessed
  emission in the Unified AGN model, selection on the basis of
  IR-luminosity will select all bolometrically luminous AGNs with $L_X
  \goa 10^{41} \ergps$. However, we note that it will also select
  dust-rich star-forming systems which may dominate the bolometric
  luminosity of the galaxy.} A particular advantage to a
volume-limited sample is that, unlike magnitude-limited surveys, they
do not suffer from radial selection effects and thus can be used to
construct volume-averaged statistics. By contrast, volume-limited
samples yield limited source number statistics due to their inability
to probe both the faintest and most luminous systems.

Given the relatively small distance scale considered here, it is
prudent to note the importance of distance measurement. For our
consider sample, luminosity distances were calculated using the cosmic
attractor model of \citet{mould00} which accurately adjusts
heliocentric redshifts to the centroid of the local group. Hence,
unlike many other local surveys, to $D<15$~Mpc our sample does not
include galaxies from local over densities such as the Virgo cluster
at $D \sim 16$~Mpc (see Appendix A1 for a detailed analysis and
validation of the considered space-volume in this survey).

The \spitzer-IRS spectroscopic data presented in GA09 was reduced
using a custom {\sc idl} pipeline which utilises the \spitzer
\ Science Center data-processing packages {\small SPICE}, {\small
  IRSCLEAN} and {\small CUBISM}. For further detailed information on
the reduction processes and spectral analyses see \S\S 2.1 and 2.2 of
GA09 and references there-in.

\subsection{[NeV] as an unambiguous AGN indicator}

Due to the very high-ionization potential of [NeV] (97.1 eV) we
consider its detection coincident with the galactic nucleus in mid-IR
spectroscopy to be an almost unambiguous identifier of AGN
activity. Theoretically, \citet{schaerer99} have predicted that
extremely hot O and B stars, in particular dense populations of
Wolf-Rayet stars, may produce ionization spectra capable of exciting
lines such as [OIV] (54.9 eV) and [NeV]. However, observationally,
GA09 found that even in extreme Wolf-Rayet galaxies, whilst [OIV] is
clearly detected in these types of systems, [NeV] emission remains
absent to the detection limits of this survey. Complimentary to this,
\citet{hao09} find from a \spitzer-IRS study of 12 Blue Compact Dwarf
galaxies that the mid-IR spectroscopy for eight of their sample
contain [OIV] emission, however none appear to be producing [NeV].

Similarly, extreme starburst driven shocks have also been predicted to
excite some high-ionization lines such as [NeV] \citep{allen08},
however these require exceptionally high velocities, and based on the
[NeV]$\lambda 14.32 \um$--[NeII]$\lambda 12.81 \um$ and
[NeIII]$\lambda 15.51 \um$--[NeII]$\lambda 12.81 \um$ emission line
ratios presented in GA09, the AGNs in this study are not consistent
with shock models. Furthermore, AGNs which contain strong
star-formation contributions to their bolometric luminosity are often
found to have relatively low mid-IR [NeV]--[NeII] and [NeIII]--[NeII]
ratios (\citealt{armus06}; Satyapal et~al. 2008; \citealt{GA09};
\citealt{dale09}). However, these ratios are not necessarily strong
tracers of the intrinsic power of the AGN. Indeed, many Seyfert 2s are
found to be hosted in galaxies where-by star-formation dominates the
IR spectral energy distribution (\citealt{weedman05};
\citealt{buchanan06}; \citealt{deo07}), thus yielding a low
        [NeV]--[NeII] ratio (log-average $\approx 0.02$); however, the
        central source may still be extremely luminous at other
        energies e.g., NGC~4945 is the most luminous local AGN at
        $E>20$~keV (Done et~al. 1996) and by contrast has a
        [NeV]--[NeII] ratio of $\approx 0.01$.

The mid-IR spectra of seventeen of the sixty-four galaxies ($\approx
27^{+8}_{-6}$ percent) presented in GA09 were found to contain the
\nev emission line, and hence, host AGN activity. These seventeen
sources are the main focus of the current paper. See Table~1.

\section{Black Hole Mass Determination}

\subsection{Archival Data}

To accurately determine the relative mass accretion rates and space
density of active SMBHs in the local Universe requires reliable SMBH
mass ($\Mbh$) estimates for the seventeen AGNs within our
volume-limited sample. Here we outline the construction of the
heterogeneous database of the most reliable available SMBH masses for
these AGNs, derived from a variety of archival sources and $\Mbh$
estimation methods.

Many of the sources in the sample are late-type galaxies hosting
relatively small bulges, and hence low-mass SMBHs. In such systems,
SMBH mass estimates are often challenging to determine as: 1)
characteristically low velocity dispersions can be difficult to
measure as they are often at the resolution limit of published
observations, 2) modeling of the contamination from composite stellar
populations can often lead to inconsistencies between published
measurements and 3) SMBH mass relations are poorly constrained at
$\Mbh \sim 10^6 \Msun$.

Given the varying degrees of accuracy associated with $\Mbh$
measurements estimated from differing methodologies, we have chosen to
prioritise the archival data (which we further expand on here) based
upon two broad categories: (i) physically motivated AGN modeling
techniques (i.e., reverberation mapping, water maser mapping and gas
kinematics of the central engine) and (ii) indirect estimations from
observational relations (the M--$\sigma_*$ and $\Mbh$--$\Lbul$
relations). The adopted $\Mbh$ for the sources and their associated
measurement methodologies are given in columns 12 and 13 of
Table~1. The database contains four SMBH measurements determined from
physically motivated AGN modeling and 13 from indirect methods.

\begin{figure}
\begin{center}
\vspace{-1.0cm}
\includegraphics[width=1.10\linewidth]{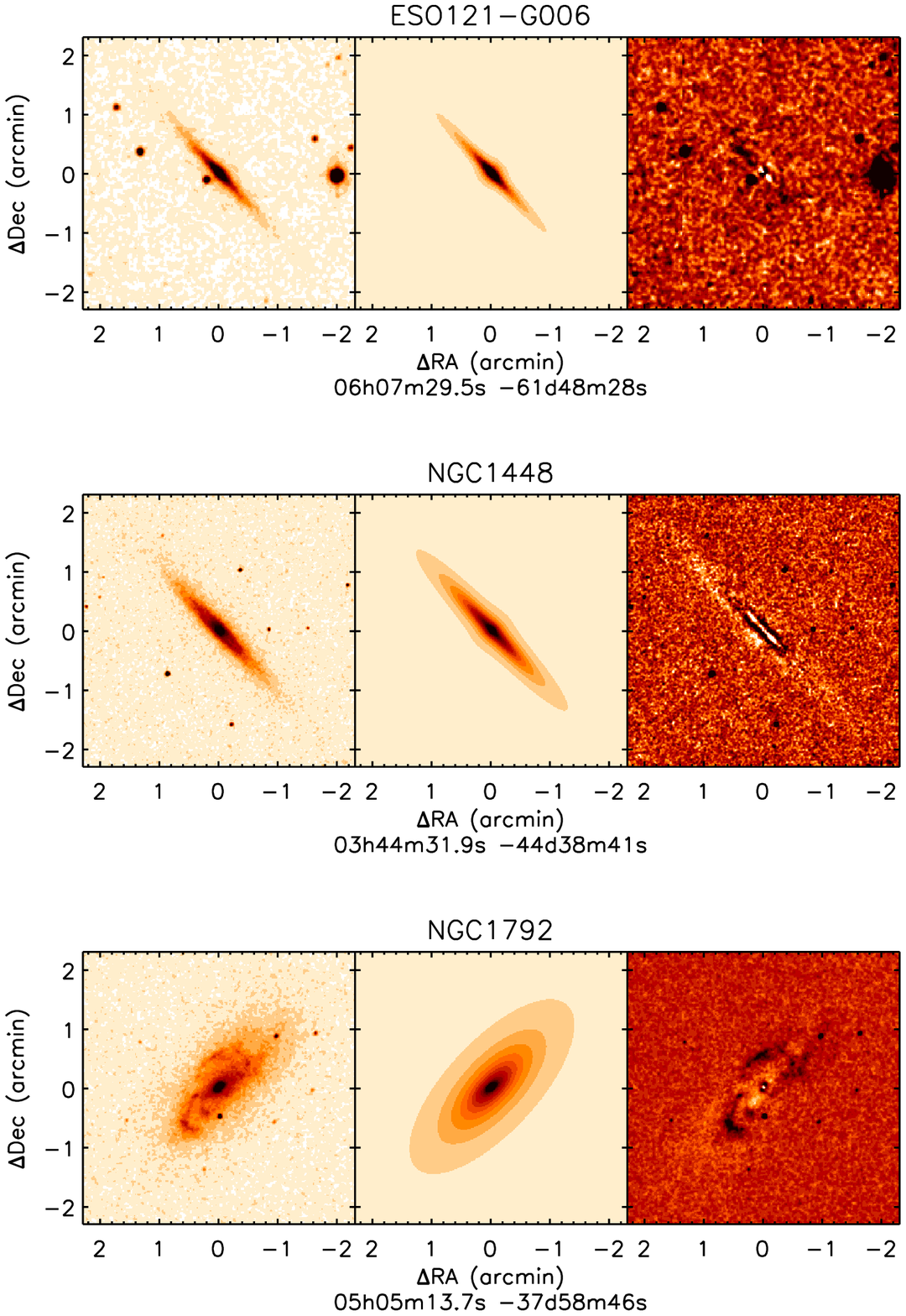}
\vspace{-2.0cm}
\caption{{\small GALFIT} \citep{galfit} two-dimensional bulge/disc
  decompositions for the three AGNs (ESO121-G006, NGC~1448 and
  NGC~1792) with $\Mbh$ estimated from the $\Mbh$--L$_{\rm K,bul}$
  relation. The panels show (left) {\it K}-band Two Micron All-Sky
  Survey (2MASS) extended source image of the galaxy, (middle) {\small
    GALFIT} model produced by fitting a Sersic profile, an exponential
  disc and a mean averaged background to the observed {\it K}-band
  image, (right) the residual image resulting from the subtraction of
  the model profile from the observed galaxy (i.e., ${\rm
    observation}-{\rm model}$). See \S 3.2.1 for a brief explanation
  of the residual images.}
\label{bd_decomp}
\end{center}
\end{figure}

\subsubsection{Direct SMBH mass constraints from reverberation mapping, maser mapping and gas kinematics}

Under the assumption that the gas in the broad-line region (BLR) is
virialised by the SMBH and the orbital motion of the gas is Keplerian,
$\Mbh$ estimations are possible through reverberation mapping
techniques \citep{blandford82}. The time-lag between changes in the
AGN continuum flux and the response of the broad-line region is used
to directly infer the size of the virial radius and hence the mass of
the SMBH (for an in-depth review see Peterson 2001). To date,
reverberation mapping is widely accepted to be the most reliable of
$\Mbh$ estimation methodologies (e.g., \citealt{wandel99b};
\citealt{peterson99}; \citealt{onken03}; \citealt{bentz09a}). Only a
minority of the galaxies in the sample (NGC~1068, NGC~4051 and
NGC~5033) are known to have detected BLRs. However, the BLRs of
NGC~1068 and NGC~5033 (both are Seyfert 1.9) are extremely weak and
are therefore likely to suffer from optical extinction. Thus, $\Mbh$
estimates using the BLR are not possible for these two AGNs. Hence,
only one object (NGC~4051; \citealt{wandel99a}) in our sample has
archival reverberation mapping data giving $\Mbh \approx
1.4^{+1.5}_{-0.9} \times 10^6 \Msun$.

Complimentary to reverberation mapping, and also assuming Keplerian
motion, mapping of water maser spots and observations of gas
kinematics within the gravitational sphere of influence of the SMBH
are thought to yield relatively precise measurements of $\Mbh$. Again,
due to the nature of the observations and the requirement of a
suitable gas disc (i.e., a relatively face-on inclination to the
observer), few $\Mbh$ estimations using these methods exist in the
current literature. Indeed, in our sample there are currently only two
AGNs (NGC~1068 and NGC~4945) with $\Mbh$ estimations from the mapping
of water maser spots ($\Mbh \approx 1.6 \times 10^7$ and $1.1 \times
10^6 \Msun$, respectively) and one AGN (NGC~5128) with a spatially
resolved gas dynamical $\Mbh$ estimate ($\Mbh \approx 2.4 \times 10^8
\Msun$).

\subsubsection{Indirect SMBH mass constraints from the M--$\sigma_*$ relation}

Since the seminal discovery that the mass of the stellar spheroid is
closely related to $\Mbh$ (Magorrian et~al. 1998), indirect $\Mbh$
estimation methods have become ubiquitous in the current literature
(e.g., \citealt{ferrarese00}; \citealt{gebhardt00};
\citealt{nelson04}; \citealt{onken04};
\citealt{gre_Ho06}). Comparisons of spatially resolved stellar
dynamics with reverberation-based $\Mbh$ estimates (over the range of
$\Mbh \approx 2 \times 10^6$--$10^9 \Msun$) show that these are
well-correlated with the effective stellar velocity dispersions
($\sigma_*$) within the galaxy bulge (i.e., the M--$\sigma_*$
relation; \citealt{gebhardt00}),
\begin{equation}
\Mbh=(1.2 \pm 0.2) \times 10^8 \Msun \left( \frac{\sigma_*}{200 \kmps} \right)^{(3.75 \pm 0.30)}  
\label{eqn:m_sig}
\end{equation}
\noindent with an intrinsic scatter of only 0.34 dex. Furthermore,
with the inclusion of a sample of dwarf Seyfert 1 galaxies, the
correlation may be reasonably extrapolated to SMBH masses in the range
$10^5$--$10^6 \Msun$ \citep{barth05}. However, we acknowledge that
to-date, the M--$\sigma_*$ relation is still poorly constrained for
$\Mbh < 10^6 \Msun$ and suffers from systematic
uncertainties. Specifically, these are caused by the lack of
homogeneous source statistics in this mass region and also the poor
understanding of the physical nature of the central region in very
late-type (Scd--Sm) galaxies (e.g., the existence of either a pseudo-
or classical bulge; \citealt{kormendy04}). Of the three AGNs in our
sample which are classified as Hubble-type Scd or later, only NGC~3621
(Sd) has an $\Mbh$ measurement estimated from the M--$\sigma_*$
relation.\footnote{The $\Mbh$ measurement for NGC~1448 (Scd) is
  estimated from the $\Mbh$--L$_{\rm K,bul}$ relation (see \S3.2.1)
  due to the lack of sufficient spectroscopic data to measure a
  stellar velocity dispersion. NGC~4945 (Scd) is an H$_2$O megamaser
  source, and hence has an accurate and direct measurement of $\Mbh$
  (Greenhill et~al. 1997). See also Table 1.}  Specifically, using the
Echellette spectrograph on Keck-II, Barth et~al. (2009) measure an
accurate line-of-sight stellar velocity dispersion of the Calcium
{\small II} triplet lines (CaT; $\lambda \lambda 8498$, 8542, 8662)
observed in the central nuclear star cluster. They find a velocity
dispersion consistent with a SMBH mass of $\Mbh \approx 3 \times 10^6
\Mbh$. We find that all of the AGNs in our sample host SMBHs with
$\Mbh \goa 10^6 \Msun$ and thus are unlikely to suffer significant
systematic uncertainities arrising from the use of the M--$\sigma_*$
relation to estimate $\Mbh$ even in the most late-type galaxies.

Ten of the 13 AGNs in our sample without direct $\Mbh$ measurements
have published $\Mbh$ estimates using the M--$\sigma_*$ relation. For
consistency purposes (and where possible) we have used the central
stellar velocity dispersions given in the recently published catalogue
of \citet{ho09}. They measure the central $\sigma_*$ for the 486
galaxies within the Palomar Survey (\citealt{ho97a,ho97b}) using the
averaged values derived from the fitting of stellar absorption
templates \citep{valdes04} to the blue (4230--5110 \AA) and red
(6210--6860 \AA) spectral ranges (i.e., the published spectroscopy
does not include standard velocity dispersion measurement features;
e.g., CaT). Where available, \citet{ho09} compare $\sigma_*$ values
derived from the modeling of the stellar absorption features to
previously published measurements from CaT lines which are available
in the HyperLeda database.\footnote{The HyperLeda database is a
  continuously updated electronic catalogue of galactic measurements
  available at http://leda.univ-lyon1.fr/. Specifically, HyperLeda
  contains a consolidated list of archival velocity dispersions for
  many nearby galaxies.} Measured errors are compared between the
Palomar $\sigma_*$ measurements and the weighted average adopted by
HyperLeda for the available published $\sigma_*$ measurements. The
final adopted measurement of $\sigma_*$ by Ho et~al. (2009) is that
with the smallest overall error. Where the values assumed by Ho
et~al. (2009) are previously published or are from HyperLeda we quote
these references in column 14 of Table 1 (6 objects). For the AGNs in
our sample which are not part of the Palomar survey, values of
$\sigma_*$ derived from direct fitting analyses of the CaT lines are
adopted from other published sources (see column 14 of Table 1; 4
objects). All final adopted $\sigma_*$ measurements are converted to
$\Mbh$ estimates using Equation 1.\footnote{We note that NGC~6300
  currently has two measurements of $\sigma_*$ from fitting of the CaT
  lines \citep{garcia05}, which were obtained through direct-fitting
  and cross-correlation analyses. For consistency with other
  measurements, we adopt the direct-fitting value of $\sigma_*$. We
  find that the derived $\Mbh$ from the cross-correlation method is a
  factor two larger; however, using this larger $\Mbh$ measurement
  will have little impact on our overall results.}

\begin{table*}
\begin{minipage}{176mm}
\begin{center}
\setlength{\tabcolsep}{0.65mm}
\caption{Catalogue of $D<15$ Mpc mid-infrared identified AGNs and derived quantities.}
\begin{tabular}{lccccccccccccc}
\hline
  \multicolumn{1}{|c|}{Common} &
  \multicolumn{1}{c|}{$D$} &
  \multicolumn{1}{c|}{Hubble} &
  \multicolumn{1}{c|}{AGN} &
  \multicolumn{1}{c|}{log(${\rm L}_{\rm [OIV]}$)} &
  \multicolumn{1}{c|}{log(${\rm L}_{\rm X}$)} &
  \multicolumn{1}{c|}{Ref.} &
  \multicolumn{1}{c|}{log(${\rm L}_{\rm bol,[OIV]}$)} &
  \multicolumn{1}{c|}{log(${\rm L}_{\rm bol,X}$)} &
  \multicolumn{1}{c|}{$K_{\rm Tot}$} &
  \multicolumn{1}{c|}{$K_{\rm Bul}$} &
  \multicolumn{1}{c|}{log($\Mbh$)} &
  \multicolumn{1}{c|}{Method} &
  \multicolumn{1}{c|}{Ref.} \\
  \multicolumn{1}{|c|}{Name} &
  \multicolumn{1}{c|}{(Mpc)} &
  \multicolumn{1}{c|}{Type} &
  \multicolumn{1}{c|}{ID} &
  \multicolumn{1}{c|}{($\ergps$)} &
  \multicolumn{1}{c|}{($\ergps$)} &
  \multicolumn{1}{c|}{${\rm L}_{\rm X}$} &
  \multicolumn{1}{c|}{($\ergps$)} &
  \multicolumn{1}{c|}{($\ergps$)} &
  \multicolumn{1}{c|}{(mag)} &
  \multicolumn{1}{c|}{(mag)} &
  \multicolumn{1}{c|}{($\Msun$)} &
  \multicolumn{1}{c|}{} &
  \multicolumn{1}{c|}{$\Mbh$} \\
  \multicolumn{1}{|c|}{(1)} &
  \multicolumn{1}{c|}{(2)} &
  \multicolumn{1}{c|}{(3)} &
  \multicolumn{1}{c|}{(4)} &
  \multicolumn{1}{c|}{(5)} &
  \multicolumn{1}{c|}{(6)} &
  \multicolumn{1}{c|}{(7)} &
  \multicolumn{1}{c|}{(8)} &
  \multicolumn{1}{c|}{(9)} &
  \multicolumn{1}{c|}{(10)} &
  \multicolumn{1}{c|}{(11)} &
  \multicolumn{1}{c|}{(12)} &
  \multicolumn{1}{c|}{(13)} &
  \multicolumn{1}{c|}{(14)} \\
\hline
 E121-G006 & 14.5 & Sc   & IR         & 39.04 & -         & -         & 41.82 & -     & 8.98 & 10.91 & $6.10^{+0.11}_{-0.51}$ & L$_{\rm Bul}$ & 14 \\
 NGC~0613  & 15.0 & Sbc  & IR         & 39.38 & -         & -         & 42.26 & -     & 7.03 & -     & $7.34^{+0.08}_{-0.15}$ & M-$\sigma_*$ & 15 \\
 NGC~0660  & 12.3 & Sa   & IR         & 39.71 & -         & -         & 42.69 & -     & 7.34 & -     & $7.35^{+0.08}_{-0.16}$ & M-$\sigma_*$ & 21 \\
 NGC~1068  & 13.7 & Sb   & IR,O,X     & 41.66 & 43.48     & 1,9,2     & 45.26 & 44.85 & 5.79 & -     & $7.20^{+0.12}_{-0.12}$ & M            & 16 \\
 NGC~1448  & 11.5 & Scd  & IR         & 39.40 & -         & -         & 42.28 & -     & 7.66 & 10.64 & $5.99^{+0.11}_{-0.52}$ & L$_{\rm Bul}$ & 14 \\
 NGC~1792  & 12.5 & Sbc  & IR         & 38.26 & -         & -         & 40.49 & -     & 7.01 & 9.08  & $6.83^{+0.12}_{-0.53}$ & L$_{\rm Bul}$ & 14 \\
 NGC~3621  & 6.6  & Sd   & IR,O       & 38.18 & -         & -         & 40.68 & -     & 6.60 & -     & $6.50^{+0.13}_{-0.27}$ & M-$\sigma_*$ & 22 \\
 NGC~3627  & 10.0 & Sb   & IR         & 38.38 & -         & -         & 40.95 & -     & 5.99 & -     & $7.30^{+0.10}_{-0.19}$ & M-$\sigma_*$ & 21 \\
 NGC~3628  & 10.0 & Sb   & IR         & 38.81 & -         & -         & 41.51 & -     & 6.07 & -     & $6.53^{+0.07}_{-0.12}$ & M-$\sigma_*$ & 20 \\
 NGC~4051  & 13.1 & Sbc  & IR,O,X     & 39.88 & 41.72     & 3,2,4,5   & 42.91 & 42.71 & 7.67 & -     & $6.15^{+0.16}_{-0.22}$ & R            & 19 \\
 NGC~4945  & 3.9  & Scd  & IR,X       & 38.72 & 42.49     & 6,2,7,4,5 & 41.40 & 43.61 & 5.23 & -     & $6.04^{+0.05}_{-0.05}$ & M            & 17 \\
 NGC~5033  & 13.8 & Sc   & IR,O,X     & 39.08 & 40.85     & 8,9       & 41.86 & 41.73 & 6.96 & -     & $7.62^{+0.09}_{-0.16}$ & M-$\sigma_*$ & 21 \\
 NGC~5128  & 4.0  & S0   & IR,X,R$^a$ & 39.38 & 41.85     & 2,9,4,5   & 42.26 & 42.86 & 3.94 & -     & $8.38^{+0.20}_{-0.26}$ & G            & 18 \\
 NGC~5194  & 8.6  & Sbc  & IR,O,X     & 38.85 & 41.11     & 10,2      & 41.56 & 42.00 & 5.92 & -     & $6.88^{+0.13}_{-0.27}$ & M-$\sigma_*$ & 15 \\
 NGC~5195  & 8.3  & Irr  & IR         & 37.89 & -         & -         & 40.30 & -     & 6.25 & -     & $7.31^{+0.07}_{-0.13}$ & M-$\sigma_*$ & 20 \\
 NGC~5643  & 13.9 & Sc   & IR,O,X     & 40.43 & 41.08     & 11,12,2   & 43.63 & 41.98 & 7.17 & -     & $6.44^{+0.11}_{-0.21}$ & M-$\sigma_*$ & 23 \\
 NGC~6300  & 13.1 & Sb   & IR,O,X     & 39.78 & 41.63     & 13,4,5    & 42.79 & 42.60 & 6.93 & -     & $6.80^{+0.11}_{-0.22}$ & M-$\sigma_*$ & 24 \\
  \hline\end{tabular}
\end{center}
{\tiny NOTES:} (1) Common galaxy name.  (2) Distance to source in
megaparsecs from the Revised Bright Galaxy Survey (RBGS; Sanders et
al. 2003). (3) Morphological type from RC3 \citep{rc3}. (4) Waveband
of AGN identification; IR: Mid-Infrared spectroscopy (GA09); O:
Optical spectroscopy (references presented in GA09); X: X-ray
spectroscopy (2--10 keV; see column 7 for references); R: Radio
observations. (5) Logarithm of [OIV] $\lambda 25.89 \um$ luminosity in
$\ergps$ calculated using [OIV] flux presented in GA09; mean
uncertainty is approximately 10 percent. (6) Logarithm of absorption
corrected hard X-ray luminosity (2--10 keV) in $\ergps$ which have
been converted to the distances given in column 2. (7) Reference for
X-ray data. (8) Logarithm of bolometric luminosity of the AGN
estimated from $L_{\rm [OIV]}$ using Equation 4. (9) Logarithm of
bolometric luminosity of the AGN estimated from $L_{\rm X, 2-10keV}$
using the bolometric corrections described in Marconi
et~al. (2004). (10) Total {\it K}-Band magnitude from 2MASS Large
Galaxy Atlas (Jarrett et~al. 2003). (11) {\it K}-Band magnitude of
bulge produced using {\footnotesize GALFIT} (\citealt{galfit}; see
section 3.2.1). (12) Logarithm of estimated black hole mass and
associated 1-sigma errors in solar masses. (13) Method of $\Mbh$
measurement; M: Maser Mapping; G: Gas Kinematics; R: Reverberation
Mapping; M-$\sigma_*$: Mass--Velocity Dispersion Correlation; L$_{\rm
  Bul}$: $K$-band Luminosity--Bulge Correlation. (14) Reference for
$\Mbh$ measurement.

\smallskip
{\tiny REFERENCES:}
(1) \citet{matt97}; (2) \citet{dadina07};
(3) \citet{pounds04}; (4) \citet{tueller08};
(5) \citet{winter09}; (6) \citet{guainazzi00};
(7) \citet{itoh08}; (8) \citet{cappi06};
(9) \citet{bird07}; (10) \citet{fukazawa01};
(11) \citet{maiolino98}; (12) \citet{guainazzi04};
(13) \citet{matsumoto04}; (14) This Paper;
(15) HyperLeda; (16) \citet{greenhill96};
(17) \citet{greenhill97}; (18) \citet{marconi01};
(19) \citet{wandel99a}; (20) \citet{ho09};
(21) \citet{barth02}; (22) \citet{barth09};
(23) \citet{whittle92}; (24) \citet{garcia05};
$^a$ for a review see \citet{israel98}.

\end{minipage}
\end{table*}

\subsection{Galaxy Decompositions using GALFIT}
\subsubsection{Indirect SMBH mass constraints from the $\Mbh$--L$_{\rm K,bul}$ relation}

Three of the AGNs within our sample (ESO121-G006, NGC~1448 and
NGC~1792) currently lack archival direct or indirect SMBH mass
constraints. Hence, for these three objects, we follow the formalism of
Marconi \& Hunt (2003; hereafter, MH03) and use 2MASS $K$-band imaging
and {\sc galfit}, the two-dimensional imaging analysis software of
\citet{galfit}, to constrain the bulge luminosities and therefore,
$\Mbh$ for these three AGNs.

Near-IR (0.9--$4.8\um$) emission is a strong tracer of stellar mass
and is less susceptible to the effects of dust/gas extinction than
optical emission. As a result of this, the {\it K}-band ($2.2\um$) is
shown to provide the strongest correlation of all near-IR bands
between the luminosity of the bulge and $\Mbh$ (MH03).

For the bulge-disc decomposition image analysis, we have obtained
archival {\it K}-band imaging for ESO~121-G006, NGC~1448 and
NGC~1792. These images were retrieved from the Two Micron All-Sky
Survey (2MASS) extended source catalogue and consist of pre-mosaicked
(1 arcsecond per pixel resolution) all-sky atlas images. The {\it
  K}-band images of the three galaxies were modeled with a central
point spread function (PSF) and a constant sky background
contribution, whilst the bulge and host-galaxy components were modeled
using variations of the Sersic profile:
\begin{equation}
\Sigma (r) = \Sigma_e e^{-\kappa[(r/r_e)^{1/n}-1]}
\label{eqn:sersic}
\end{equation}
\noindent where $r_e$ is the effective radius of the profile,
$\Sigma_e$ is the surface brightness at the effective radius, $n$ is
the power-law (Sersic) index, and $\kappa$ is coupled to $n$ such that
half of the total flux of the object is within the effective
radius. We employ two special forms of the Sersic profile in our {\sc
  galfit} modeling, the exponential ($n=1$) and the de Vaucouleurs
($n=4$) profiles, which are classically used to model galactic discs
and bulges, respectively.

\citet{haussler07} have shown that the reliability of the fitting
parameters produced by {\sc galfit} are strongly dependent on the
initial estimates. {\sc galfit} will, in general, fail to find the
overall global chi-squared ($\chi^2$) minimum to the fit if the
initial estimates are poorly constrained. Thus, to reduce this
systematic effect, and aid the fitting routine, we use a simplified
1-dimensional fit to produce initial estimates of the fitting
parameters. A 1-dimensional surface-brightness slice of the {\it
  K}-band image was taken across the major axis of each of the
galaxies. A surface brightness profile extending from the nucleus was
produced by averaging the two semi-major axes from the slice, and
removing the measured background flux. Few spiral galaxies are found
to host bulges with true de Vaucouleurs profiles, thus a global
$\chi^2$ reduction process was used to simultaneously fit a
generalised Sersic profile and a fixed ($n=1$) exponential disc to the
1-d surface brightness profile. From these, we calculate Sersic and
disc radii, as well as the Sersic index of the bulge. Combining the
1-d parameter estimates with the total {\it K}-band magnitude from the
2MASS Large Galaxy Atlas \citep{jarrett03}, we generate an appropriate
set of constraints and initial parameters to be input to {\sc galfit}.

Using the derived parameter estimates, {\sc galfit} is used to fit a
generalised 2-d Sersic profile with an exponential disc to the {\it
  K}-band image. To again aid the {\sc galfit} reduction analysis,
particular attention is paid to simulating accurate PSFs for the 2MASS
images using known standard stars (J.~R.~Lucey private
communication).\footnote{A detailed discussion of {\sc galfit}
  problems caused by poor PSF modeling can be found in
  \citet{bentz09b}} Results of this bulge/disc reduction for the three
objects are presented in Fig.~\ref{bd_decomp} and column 11 of
Table~1.

We have directly tested our robust {\sc galfit} method using the
late-type galaxies (i.e., S0 or later) presented in the dataset of
MH03 and find close agreement ($\approx 0.1$ dex). We do note however,
that we find a systematic offset of a factor $\approx 2$ in bulge
luminosity for the AGNs in MH03 that are hosted in low-inclination
angle late-type galaxies, which is likely to be caused by {\sc galfit}
over-estimating the contribution of the bulge to the total flux of the
galaxy. Indeed, when directly comparing a sample of reverberation
mapped X-ray detected AGNs to $\Mbh$ estimations using the MH03
formalism, \citet{vasudevan09a} find similar results. However, the
three galaxies fitted in our sample are all moderately to
highly-inclined and thus this systematic effect will be negligible.

In Fig. \ref{bd_decomp} we show the three {\sc galfit} produced image
cubes obtained following our bulge/disc fitting routines. Within each
of the residual (observed -- model) images it is clear that the bulge
is well fitted by a Sersic profile. The edge-on galaxy ESO121-G006 is
well fit by an exponential disc combined with a Sersic profile, with
no distinguishing residual features. The residual of NGC~1448
highlights the existence of its spiral arms and shows the presence of
a truncated disc combined with a possible bar structure which our
simplified modeling technique is incapable of fitting; however, the
bulge fit does not appear to be compromised. Indeed, our derived
$\Mbh$ estimations for ESO121-G006 and NGC~1448 are consistent with
the $\Mbh$ upper limits obtained from stellar mass-to-light ratio
analyses \citep{ratnam00}. The residual image of the moderately inclined
galaxy, NGC~1792, contains strong spiral arms as well as a point-like
nuclear source. We note that due to the inclination angle of this
source, the derived $\Mbh$ may be systematically over-estimated by a
factor of $\approx 2$ (see above). Thus, our derived Eddington rate
for NGC~1792, presented in \S5.1, should be considered a lower limit.

From the obtained {\it K}-band bulge magnitudes (column 11, Table 1) we
calculate bulge luminosities ($L_{\rm K,bul}$). Using the well
established $\Mbh$ to bulge {\it K}-band luminosity relation of MH03
(hereafter, $\Mbh$--$L_{\rm K,bul}$),
\begin{equation}
{\rm log \ } \Mbh = (8.08 \pm 0.10) + (1.21 \pm 0.13) {\rm log} \left( \frac {L_{\rm K,bul}}{8 \times 10^{10}} \right)
\label{eqn:mh03}
\end{equation}
\noindent we calculate $\Mbh$ for ESO~121-G006, NGC~1448 and
NGC~1792. The 1-$\sigma$ uncertainties associated with the $\Mbh$
values are assessed by combining in quadrature the bulge magnitude
errors as calculated by {\sc galfit} with the intrinsic dispersion
observed in the $\Mbh$--L$_{\rm K,bul}$ relation. The resultant $\Mbh$
estimates and 1-$\sigma$ uncertainties are provided in column 12 of
Table 1.

\section{Bolometric Corrections}

To determine the relative mass accretion rates, and hence the average
growth times of SMBHs in the local Universe we require relatively
accurate estimates of the bolometric luminosities of the AGNs ($L_{\rm
  Bol,AGN}$) in our sample. We use two methods to estimate $L_{\rm
  Bol,AGN}$: 1) a direct approach using the best available measured
hard X-ray (2--10 keV) luminosities, and 2) a well-constrained $L_{\rm
  Bol,AGN}$--[OIV] luminosity relation to infer the intrinsic
luminosity of the AGN (e.g., \citealt{dasyra08};
\citeauthor{melendez08b} 2008, hereafter, M08).

\subsection{Hard X-ray Luminosity as a Tracer of the Bolometric Luminosity of an AGN}

High quality hard X-ray spectral analyses arguably provide the most
unambiguous method for measuring the intrinsic luminosity of an AGN
since: 1) X-rays are relatively unaffected by dust extinction; 2)
intrinsic absorption can be directly constrained from high S/N data;
3) star-formation contamination is often found to be negligible. We
therefore divide our sample into two categories based on the quality
and energy range of their available published X-ray data: 1) AGNs with
high S/N X-ray spectra where $N_{\rm H}$ has been accurately
constrained and/or AGNs with $E>10$~keV constraints where the observed
X-ray emission will only be strongly absorbed for heavily
Compton-thick ($N_{\rm H} > 10^{25} \pcmsq$) sources (8 AGNs); and 2)
those AGNs with no or low S/N X-ray data, i.e., where there are
insufficient counts to accurately determine $N_{\rm H}$ and the X-ray
flux could have large contributions from star-formation (9
AGNs). Hence, we specifically do not estimate $L_{\rm Bol,AGN}$ for
those AGNs with $L_X$ measurements using {\it Chandra} that also have
no further hard X-ray spectral constraints ($E > 10$~keV) due to the
limited band-pass of the instrument at $z \sim 0$ (0.5--8~keV).

For the AGNs in our sample currently with either high S/N spectroscopy
or $E>10$~keV constraints, we estimate $L_{\rm Bol,AGN}$ using the
Eddington ratio independent AGN bolometric corrections outlined in
Equation 21 of \citet{marconi04}. We note that \citet{vasudevan09b}
have suggested that the bolometric correction factor ($\kappa_{\rm
  2-10keV} = L_{\rm bol}/L_{\rm 2-10keV}$) may be a function of the
Eddington ratio of a considered source. Values of $\kappa_{\rm
  2-10keV} \sim 10$--30 are considered to be relatively low bolometric
corrections and are generally found in AGNs with $\eta < 0.1$ (e.g.,
Vasudevan \& Fabian 2009; \citealt{vasudevan10}). For the sample of
X-ray detected AGNs considered here, we calculate similarly consistent
values of $\kappa_{\rm 2-10keV} \sim 8$--30, and thus conclude that
the Eddington ratio is unlikely to be dominating the bolometric
corrections adopted here from \citet{marconi04}.

For further consistency, all archival X-ray luminosities were adjusted
to the distances adopted in column 2 of Table 1. Final adopted $L_X$
measurements, estimated $L_{\rm Bol,AGN}$ from $L_X$, and archival
references for $L_X$ are given in columns 6, 7 and 9 of Table 1,
respectively.

\subsection{[OIV] Luminosity as a Tracer of the Bolometric Luminosity of an AGN}

\begin{figure}
\includegraphics[width=1.05\linewidth]{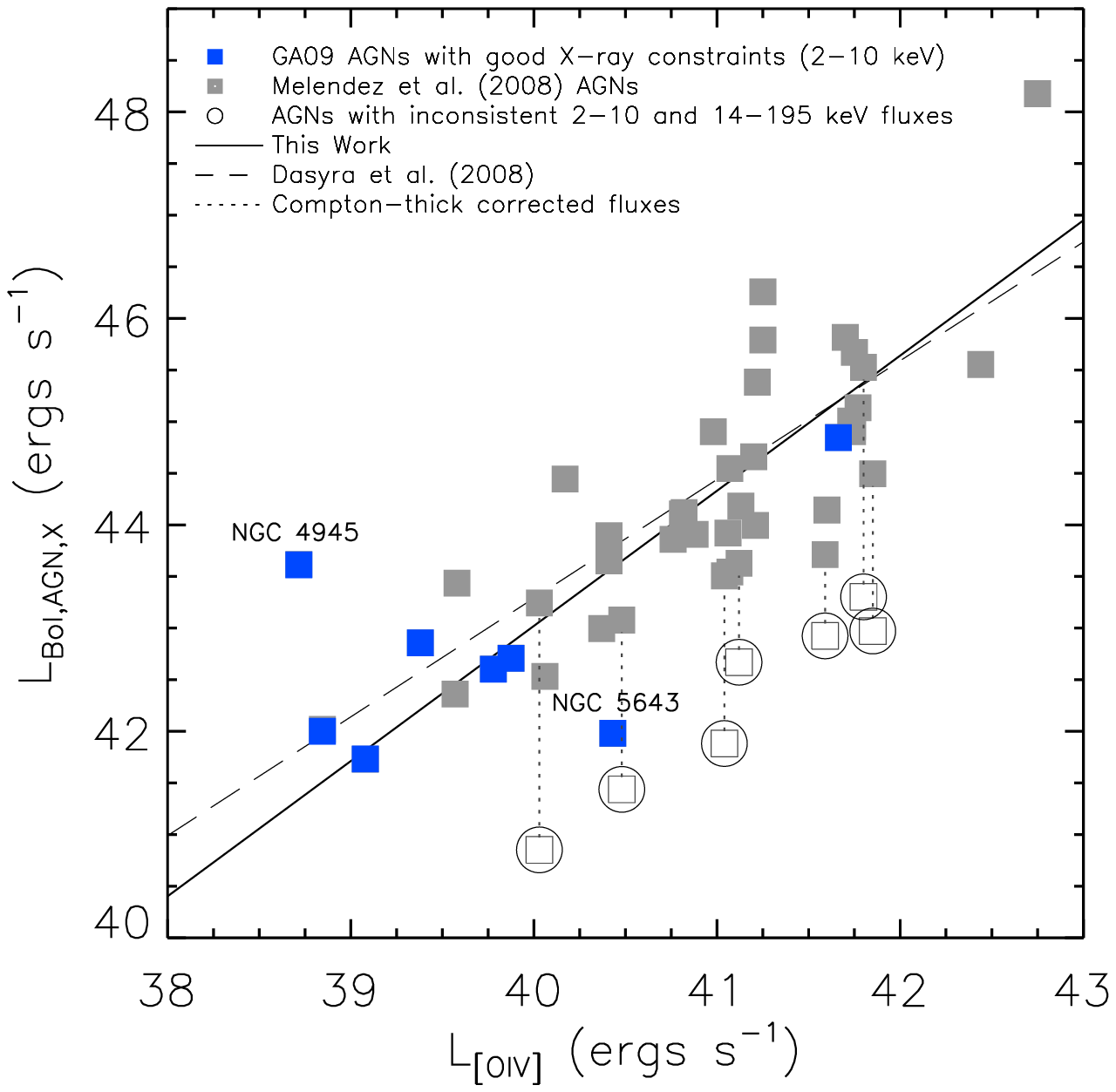}
\caption{X-ray derived AGN bolometric luminosity ($L_{\rm Bol,AGN,X}$)
  versus [OIV] $\lambda 25.89 \um$ luminosity ($L_{\rm [OIV]}$). The
  objects shown are from this work (blue squares) and those in the
  {\it Swift}-BAT survey with detected [OIV] emission (M08; grey
  squares). The solid line is a best-fit bisector and was obtained
  using the {\tiny IDL}-based {\tiny ROBUST\_LINEFIT} algorithm. The
  dashed line is the $L_{\rm Bol,AGN}$--$L_{\rm [OIV]}$ relation of
  Dasyra et al. (2008) derived from the optical luminosity at
  5100\AA. We additionally highlight those galaxies in M08 with
  inconsistent 2--10 and 14--195 keV luminosities (open
  circles). These seven galaxies and their alternatively published
  X-ray luminosities (which we adopt in our analysis; dotted line) are
  shown in Table 2; we find that five are identified as Compton thick
  AGNs in the literature and the other two are X-ray variable. The
  galaxies from our $D<15$ Mpc sample which lie significantly off the
  derived correlation are labeled (NGC~4945 and NGC~5643); see text
  for further details.}
\label{fig:loiv_lbol}
\end{figure}

For those galaxies in the sample without good-quality hard X-ray
constraints we require an additional approach to estimate the
intrinsic luminosity of the AGN, and hence the relative mass-accretion
rate. Here we build-upon an $L_{\rm Bol,AGN}$ estimation which relies
on the AGN-produced [OIV] $\lambda 25.89 \um$ luminosities ($L_{\rm
  [OIV]}$) for the AGNs in our sample (e.g., Dasyra et~al. 2008; M08).

Based on the simplest Unified Model of AGN \citep{antonucci93}, the
hot dust within the predicted torus, close to the central engine,
reprocesses absorbed UV, optical and X-ray emission into mid-IR
emission. Hence, AGN emission detected at IR wavelengths is likely to
be isotropic and independent of viewing angle. As discussed in \S2,
the detection of high-ionisation [NeV] emission (97.1 eV) coincident
with the nucleus of a galaxy is considered a robust indicator of AGN
activity (e.g., Armus et~al. 2006; GA09). Complimentary to this, GA09
find that [NeV] emission is also well correlated with [OIV] emission
(54.9~eV) with an intrinsic scatter of only 0.24 dex. As [NeV]
emission, and thus [OIV] emission, do not suffer from significant
star-formation contamination and are both comparatively
extinction-free, they may be used as relatively clean proxies for the
bolometric luminosity of the AGNs presented here ($L_{\rm
  Bol,AGN}$).\footnote{We note that since star-formation can also
  produce [OIV] emission, the [NeV]--[OIV] relation may be unreliable
  for sources with exceedingly high star-formation rates (e.g., Ultra
  Luminous Infrared Galaxies with $L_{\rm IR} > 10^{12} \ergps$;
  ULIRGs). However, there are no ULIRGs within the sample considered
  here.} Indeed, for a sample of 35 well-studied optically unobscured
AGNs, Dasyra et~al. (2008) show that both [NeV] and [OIV] emission are
well correlated with the luminosity of the 5100 \AA \ optical
continuum, and hence $L_{\rm Bol,AGN}$ with an intrinsic scatter of
0.46 and 0.47 dex, respectively. However, the relation of Dasyra
et~al. (2008) is derived from AGNs with $L_{\rm [OIV]} > 2 \times
10^{40} \ergps$ ($L_{\rm [NeV]} > 7 \times 10^{39} \ergps$), and hence
we test whether it may be reliably extrapolated to the more modest
luminosity AGNs considered here (log-average $L_{\rm [OIV]} \approx 2
\times 10^{39} \ergps$).

We combine our robustly adopted $L_{\rm Bol,AGN}$ from 2-10 keV flux
measurements with the X-ray catalogue of nearby ($z < 0.08$) Seyfert
galaxies in the {\it Swift}-BAT survey which have published [OIV]
luminosities in M08. The catalogue of sources in M08 contain 2--10~keV
luminosities obtained primarily from {\it ASCA} data, 14-195 keV
luminosities from the {\it Swift}-BAT survey and [OIV] luminosities
from {\it Spitzer}-IRS spectroscopy. Combining the M08 sample with our
8 AGNs with high-quality X-ray constraints, the range covered in
$L_{\rm [OIV]}$ is $\approx (0.7$--$7000) \times 10^{39} \ergps$. For
consistency, we convert the M08 2--10 keV luminosities to $L_{\rm
  Bol,AGN}$ using the same bolometric corrections adopted in \S4.1. In
Fig.~\ref{fig:loiv_lbol} we plot $L_{\rm Bol,AGN}$ versus $L_{\rm
  [OIV]}$ for the M08 sample and the 8 AGNs in our GA09 sample with
good-quality hard X-ray constraints (grey filled squares and blue
filled squares, respectively).

\begin{table}
\begin{center}
\setlength{\tabcolsep}{0.95mm}
\caption{Catalogue of revised 2--10 keV luminosities for a subset of M08 sample.}
\begin{tabular}{lcccc}
\hline
  \multicolumn{1}{|c|}{Common} &
  \multicolumn{1}{c|}{log(${\rm L}_{\rm 2-10keV,M08}$)} &
  \multicolumn{1}{c|}{log(${\rm L}_{\rm 2-10keV,alt.}$)} &
  \multicolumn{1}{c|}{C-thick} &
  \multicolumn{1}{c|}{Ref.} \\
  \multicolumn{1}{|c|}{Name} &
  \multicolumn{1}{c|}{($\ergps$)} &
  \multicolumn{1}{c|}{($\ergps$)} &
  \multicolumn{1}{c|}{AGN?} &
  \multicolumn{1}{c|}{} \\
  \multicolumn{1}{|c|}{(1)} &
  \multicolumn{1}{c|}{(2)} &
  \multicolumn{1}{c|}{(3)} &
  \multicolumn{1}{c|}{(4)} &
  \multicolumn{1}{c|}{(5)} \\
\hline
Circinus & 40.58 & 42.04 & $\surd$  & 1 \\
Mrk~3    & 41.95 & 43.20 & $\surd$  & 2 \\
NGC~1365 & 40.99 & 42.40 & $\surd$  & 3 \\
NGC~2992 & 41.69 & 42.50 & $\times$ & 4 \\
NGC~3079 & 40.02 & 42.18 & $\surd$  & 5 \\
NGC~4388 & 41.91 & 42.57 & $\times$ & 6 \\
NGC~6240 & 42.23 & 44.00 & $\surd$  & 7 \\
  \hline\end{tabular}
\end{center}
{\tiny NOTES:} (1) Common galaxy name.  (2) Logarithm of 2--10 keV
luminosity adopted by M08. (3) Logarithm of absorption corrected 2--10
keV luminosity adopted in this work from individual studies (2--10~keV
luminosities were adjusted using our adopted distances). (4) Is the
AGN a Compton thick source? (5) Reference for adopted $L_{\rm 2-10
  keV}$ measurement.

{\tiny REFERENCES:}
(1) \citet{yang09};
(2) \citet{awaki08};
(3) \citet{risaliti09};
(4) \citet{yaqoob07};
(5) \citet{iyomoto01};
(6) \citet{shirai08};
(7) \citet{vignati99}
\end{table}

Seven of the AGNs in M08 were found to have unusually low 2--10~keV
luminosities when compared to the quoted 14-195~keV luminosities,
which are highlighted with open circles in Fig.~\ref{fig:loiv_lbol}
and are shown in Table 2. Using high-quality X-ray spectral analyses,
in the literature we find that five of these AGNs are identified as
Compton thick AGNs (see column 4 of Table~2). This additional
obscuration does not appear to have been accounted for in the adopted
value of M08, and therefore we have chosen to select measurements of
$L_{\rm 2-10 keV}$ from the literature, which now give good agreement
with 14--195~keV luminosities (see columns 2, 3 and 5 of Table 2;
dotted-lines in Fig.~\ref{fig:loiv_lbol}). The other two AGNs
(NGC~2992 and 4388) in M08 with inconsistent 2--10 and 14--195~keV
fluxes are found to be highly variable (\citealt{beckmann07} and
\citealt{elvis04}, respectively).

With the inclusion of the obscuration-corrected 2--10~keV luminosities
to infer $L_{\rm Bol,AGN}$, we find a strong correlation between
$L_{\rm [OIV]}$ and $L_{\rm Bol,AGN}$ which is characterised by the
equation:
\begin{eqnarray}
{\rm log} \left(\frac{L_{\rm bol,AGN}}{10^{44} \ergps}\right) = (0.38 \pm 0.09) + \nonumber \\
(1.31 \pm 0.09) {\rm log} \left( \frac{L_{\rm [OIV]}}{10^{41} \ergps} \right)
\end{eqnarray}
\noindent with an intrinsic scatter in the data of $\approx
0.35$~dex. We find good agreement with the
Dasyra et~al. (2008) relation (dashed-line) in the region $L_{\rm
  [OIV]} \approx 10^{40}$--$10^{43} \ergps$ (i.e., where the Dasyra
et~al. (2008) relation is well sampled). However, for $L_{\rm [OIV]} <
10^{40} \ergps$ we show that the Dasyra et~al. (2008) relation will
over estimate $L_{\rm Bol,AGN}$ by typically $\approx 0.5$ dex.

\begin{figure*}
\begin{center}
\includegraphics[width=0.95\textwidth]{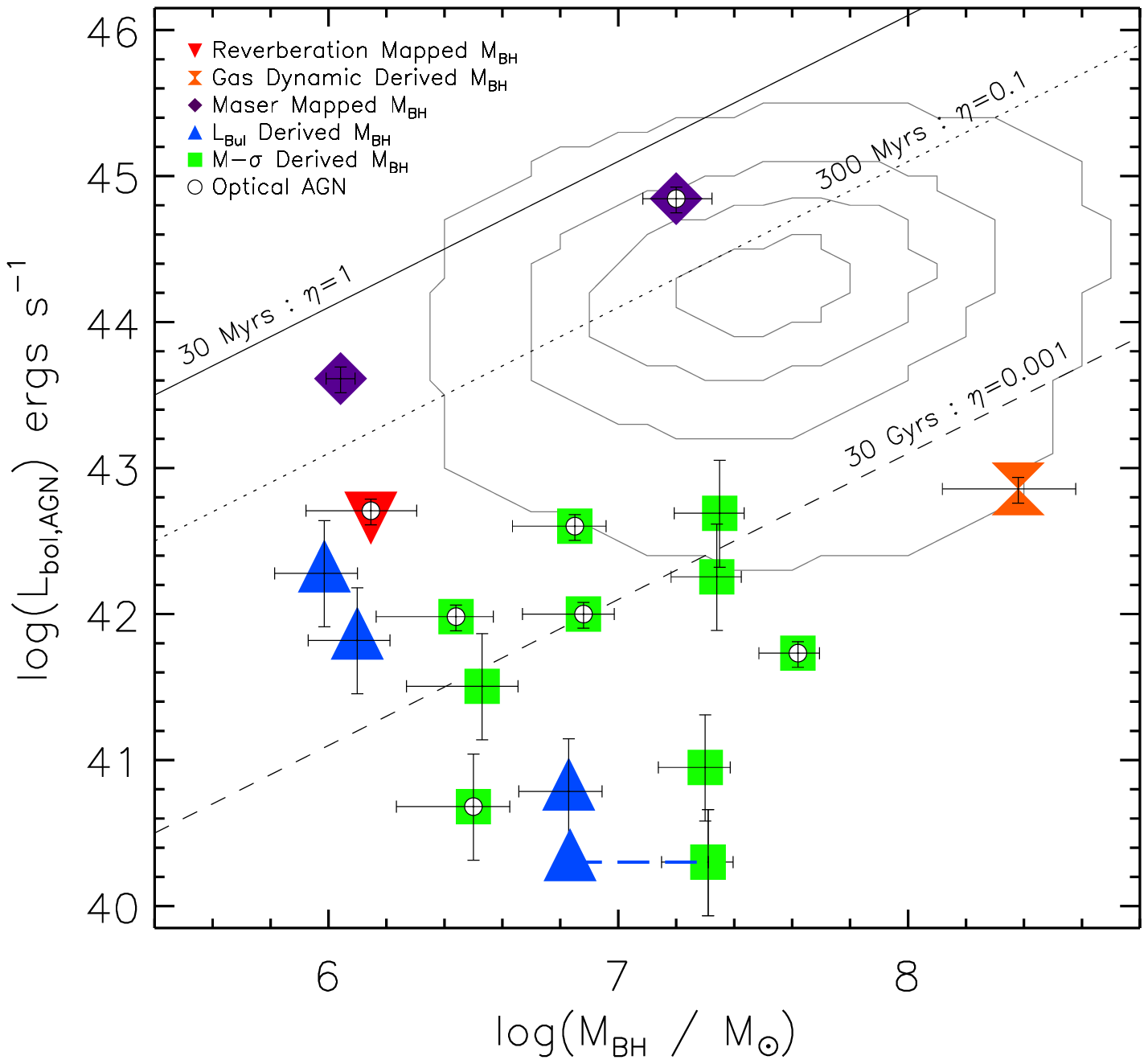}
\caption{AGN bolometric luminosity ($L_{\rm Bol,AGN}$; in $\ergps$) is
  plotted against SMBH mass ($\Mbh$) for the $D<15$~Mpc mid-infrared
  identified AGNs presented in GA09. Associated 1--$\sigma$ error bars
  for $\Mbh$ and $L_{\rm Bol,AGN}$ estimations are shown (see \S3 and
  \S5.1, respectively for details of their derivations). AGNs which
  are previously identified in optical surveys are highlighted with
  open circles. AGNs with $\Mbh$ estimates from reverberation mapping
  (downward triangles), gas dynamics (hour glass), maser mapping
  (diamond), the M--$\sigma_{*}$ relation (squares) and the
  $\Mbh$--L$_{\rm K,bul}$ relation (upward triangles) are
  plotted. NGC~5195 is represented with both an upward triangle and
  square (with a dashed-line connector) as both of the $\Mbh$
  estimates for this galaxy are highly uncertain given its irregular
  morphology. Constant ratios of Eddington luminosity and their
  implied SMBH mass-doubling times are illustrated for $\eta =
  10^{-3},10^{-1},1$ (30~Gyrs, 300~Myrs and 30~Myrs; solid line,
  short-dash and long-dash, respectively). Contours are shown for the
  active galaxies in the SDSS optical survey of H04. In general, we
  probe lower SMBH masses and AGN luminosities than those of H04, and
  we find that the majority of these AGNs would not be detected using
  optical SDSS data alone.}
\label{fig_2}
\end{center}
\end{figure*}

In Fig.~\ref{fig:loiv_lbol}, we highlight two AGNs from the $D<15$~Mpc
sample (NGC~4945 and 5643) which appear to be significant outliers of
the observed correlation. NGC~5643 possibly harbours a variable
central source. From the detection of strong Fe K$\alpha$ emission,
\citet{maiolino98} suggest from using {\it BeppoSAX} data that
NGC~5643 is possibly Compton thick (N$_{\rm H} > 10^{25} \pcmsq $);
however, Guainazzi et al. (2004) find using {\it XMM-Newton} data,
that it may be Compton thin with N$_{\rm H} \approx (6$--$10) \times
10^{23} \pcmsq$. Thus, from current available data, the true intrinsic
luminosity of the AGN is highly uncertain. Here we conservatively
adopt the Compton thin $L_X$ value of \citet{guainazzi04}; however, we
note that if we use the value of \citet{maiolino98}, then NGC~5643
would lie on our derived relationship. NGC~4945, by contrast, has
observations using {\it Ginga}, {\it ASCA}, {\it OSSE}, the {\it Rossi
  X-ray Timing Explorer}, {\it BeppoSAX}, {\it Suzaku} and most
recently {\it Swift}-BAT (\citealt{iwasawa93}, \citealt{tanaka94},
\citealt{done96}, \citealt{madejski00}, \citealt{guainazzi00},
\citealt{itoh08} and \citealt{tueller08}, respectively) all of which
provide excellent and consistent spectral constraints over a wide
X-ray band (1--200 keV) showing that the central source is Compton
thick ($N_H \approx 4 \times 10^{24} \pcmsq$), with $L_{\rm X,2-10
  keV} \approx (3$--$10) \times 10^{42} \ergps$ and an observed
intrinsic variability of a factor $\approx 2$. Here we adopt the
luminosity from the most recent observation by {\it Swift}, $L_{X}
\approx 3 \times 10^{42} \ergps$. On the basis of our mid-IR
constraints, it would therefore appear that NGC~4945 is under-luminous
in [OIV] flux by a factor of $\approx 50$ (see Table 1 and
Fig. \ref{fig:loiv_lbol}). We suggest this deficit in observed [OIV]
flux is unlikely to be due to host galaxy extinction; the required
absorption to account for a factor of $\approx 50$ flux difference is
$A_V \approx 240$~mags ($N_H \approx 5 \times 10^{23} \pcmsq$ assuming
typical dust-to-gas ratios; using $A_V/{\rm E}(B-V) = 3.1$). Another
possible explanation is a temporary decoupling of the X-ray emitting
and narrow-line regions (i.e., the highly-luminous state of NGC~4945
may be a somewhat recent event). Given the spatial difference (and
hence, the light-travel time) between the two emission regions:
$10^{-4}$~pc and 1--10~pc, respectively, the photoionisation of the
narrow-line region, and thus the observed [OIV] emission, may take
$\approx 100$ years to respond to the changes in the X-ray emitting
region. We therefore suggest that the intrinsic scatter in the
observed [OIV]--$L_{\rm bol,AGN}$ relation may be significantly
reduced if it was possible to account for variability in the central
region of all of the AGNs. Indeed, we find the average dispersion
decreases to $\approx 0.31$ dex if we remove NGC~4945 from our
analysis.

Using Equation 4 and $L_{\rm [OIV]}$ from GA09 (column 5 of Table 1),
we estimate $L_{\rm Bol,AGN}$ (column 7 of Table 1) for those AGNs in
our sample currently without good hard X-ray measurements. We use
these $L_{\rm Bol,AGN}$ estimates to assess the relative mass
accretion rates ($\eta \sim {\rm L}_{\rm Bol,AGN}/{\rm L}_{\rm Edd}$)
of the SMBHs in our $D<15$~Mpc sample.


\section{Results and Discussion}

From a volume-limited sample of 64 bolometrically luminous galaxies to
$D<15$ Mpc ($\approx 94$ percent complete) GA09 unambiguously
identified seventeen ($\approx 27^{+8}_{-6}$ per cent) sources to be
hosting AGN activity in galaxies with $\Lir > 3 \times 10^9 \Lsun$,
using [NeV] $\lambda 14.32 \um$ emission as a robust AGN
indicator. Using the SMBH mass and AGN bolometric luminosity estimates
derived here, we discuss the relative mass accretion rates of these
seventeen AGNs and use them to derive the average present-day growth
times of SMBHs in the very nearby Universe. Furthermore, we evaluate
the unique contribution that our new optically unidentified AGNs make
to the space density of active SMBHs in the local Universe, which have
until now been previously derived from large-scale optical surveys
(e.g., H04; Greene \& Ho 2007).

\subsection{Derived AGN Properties and Relative Mass Accretion Rates}

In Fig.~\ref{fig_2}, we plot $L_{\rm Bol,AGN}$ against our adopted
$\Mbh$ estimates (the associated 1-$\sigma$ errors for $\Mbh$
measurements are described in section 3) for the 17 AGNs in our
volume-limited sample. $L_{\rm Bol,AGN}$ is inferred from either
accurate intrinsic high-quality hard X-ray (2--10 keV) constraints
(where available) or AGN-produced [OIV] $\lambda 25.80 \um$ emission
(see \S4.1 and 4.2). The $L_{\rm Bol,AGN}$ 1-$\sigma$ errors for the
sources with hard X-ray constraints are the result of combining the
uncertainty in the $L_{X,2-10keV}$ measurement with that of the mean
spread in the bolometric correction factor employed from
\citet{marconi04}. For those AGNs with $L_{\rm Bol,AGN}$ derived from
$L_{\rm [OIV]}$, the error is derived from the uncertainty in $L_{\rm
  [OIV]}$ as quoted in GA09 combined in quadrature with the intrinsic
scatter of the empirical [OIV]--$L_{\rm bol,AGN}$ relation (equation
4).

Seven objects within the sample have been classified as optical AGNs
from previous surveys (see GA09 and Table~1) using typical optical
emission-line diagnostics (e.g., the Baldwin-Phillips-Terlevich
diagnostic diagrams; \citeauthor{bpt} 1981); however, all have
detected \nev emission, and thus are unambiguously identified to host
AGNs at mid-IR wavelengths (GA09).\footnote{We note that NGC~3627 is
  ambiguously classified as T2/S2 from the optical spectroscopy
  presented in \citet{ho97a}, and thus may host an active central
  source. However, \citet{roberts01} suggest from its optical
  emission-line ratios that NGC~3627 is most likely a LINER/HII
  composite.}  We find that with the exception of NGC~5128 (Centaurus
A), our sample is dominated by AGNs with SMBHs in the mass range $\Mbh
\approx (0.1$--$5) \times 10^7 \Msun$ (median of $\Mbh \approx 7
\times 10^6 \Msun$). Due to the irregular structure of one of the
galaxies in the sample (NGC~5195), $\Mbh$ is poorly determined; in
Fig.~\ref{fig_2} we plot $\Mbh$ estimates from both the M-$\sigma_*$
and $\Mbh$--L$_{\rm K,bul}$ relations (connected blue-dashed line).

\begin{figure}
\begin{center}
\includegraphics[width=1.05\linewidth]{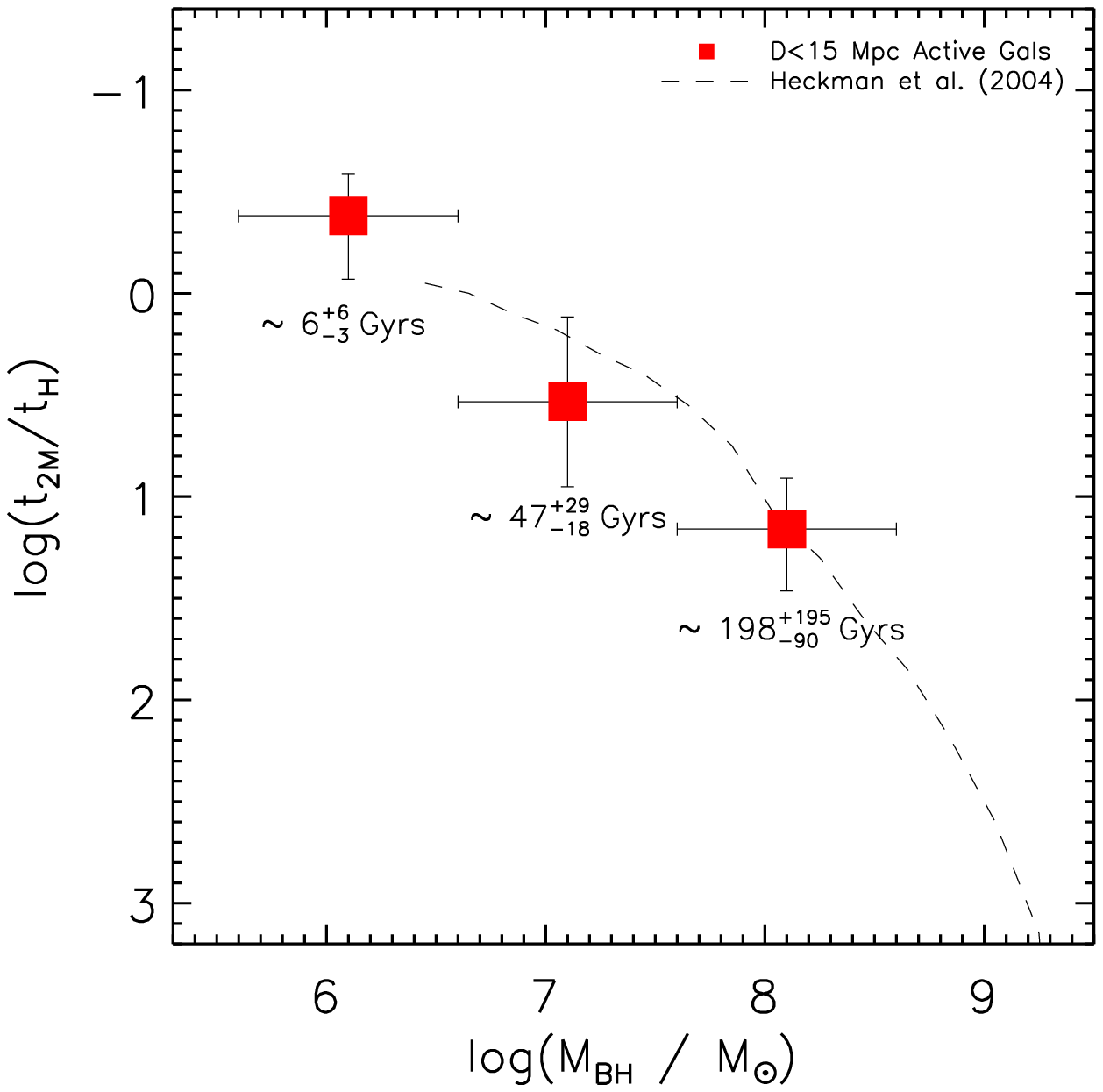}
\caption{$\Mbh$ is plotted against the characteristic mean mass
  doubling time ($t_{2M}$) of a SMBH in units of Hubble-time ($t_H$)
  for the $D<15$~Mpc AGNs. Growth time errors are calculated from the
  log-normal standard deviations of the sample. For comparison, the
  growth time function of H04 is shown (dashed line). We find good
  agreement with H04 over the region $\Mbh \approx 0.3$--$10 \times
  10^7 \Msun$ and further extend the growth time constraints to lower
  SMBH masses ($\Mbh \approx 10^6 \Msun$).}
\label{fig_4}
\end{center}
\end{figure}

We find the AGNs in our sample are spread over a wide range of
bolometric luminosities, $L_{\rm Bol,AGN} \approx 10^{40}$--$10^{45}
\ergps$. To assess the relative mass-accretion rates of the sample
($L_{\rm Bol,AGN} / L_{\rm Edd} \sim \eta$), we over-plot lines of
constant Eddington ratios ($\eta \approx 10^{-3},10^{-1},1.0$; derived
following \citealt{rees84}) and their associated mass-doubling times
($t \approx 30,0.3,0.03$~Gyrs, respectively). Given the large range in
bolometric luminosities, it is not surprising that the AGNs in the
sample are found to be accreting at rates covering over 5 orders of
magnitude ($\eta \approx 10^{-5}$--1). With the exception of a few
AGNs, the observed range in Eddington ratios is found to be roughly
consistent with those found by H04 for active galaxies (solid contours
in Fig.~\ref{fig_2}).

As our work is not limited by the spectral resolution of the SDSS
(i.e., with a limit of $\Mbh \ga 3 \times 10^6 \Msun$), we show in
Fig.~\ref{fig_2} that significant accretion, $\eta > 10^{-3}$ (i.e.,
radiatively efficient accretion systems; e.g., thin discs) occurs onto
SMBHs with $\Mbh \approx (1$--$3) \times 10^6 \Msun$. The majority of
these low-mass, rapidly-accreting SMBHs are hosted in late-type,
disc-dominated spiral galaxies (Sc--Sd). By contrast, it is generally
assumed that gas-rich late-type spirals are preferentially inactive
galaxies and that a large bulge may be a necessary component for the
existence of a SMBH, and thus a luminous AGN. Furthermore, of the four
AGNs within the sample with SMBHs consistent with $\Mbh \approx 10^6
\Msun$, we find that three sources are not identified as AGNs in
sensitive optical surveys. This indicates that significant SMBH
accretion may be missed by statistically-large optical surveys such as
H04 even if the spectral resolution was sufficient to identify SMBHs
down to $\Mbh \approx 10^6 \Msun$.

For the subset of our AGN sample which host SMBHs with $\Mbh \goa 3
\times 10^6 \Msun$, we find that many of the optically unidentified
AGNs are accreting at relatively low Eddington ratios ($\eta \loa
10^{-3}$), and are unlikely to make a significant additional
contribution to the present-day growth of SMBHs. However, these same
AGNs may form part of a separate, underlying population of radiatively
inefficient accretion systems such as advection dominated accretion
flows (ADAFs; e.g., \citealt{adaf}) or those which contain
optically-thick slim-discs. Further spectral analysis of the X-ray
data may distinguish between these particular accretion systems, but
is beyond the scope of these analyses (see Goulding et~al. in
preparation).

\subsection{The Present-Day Growth of SMBHs}

\begin{figure*}
\begin{center}
\includegraphics[width=0.95\textwidth]{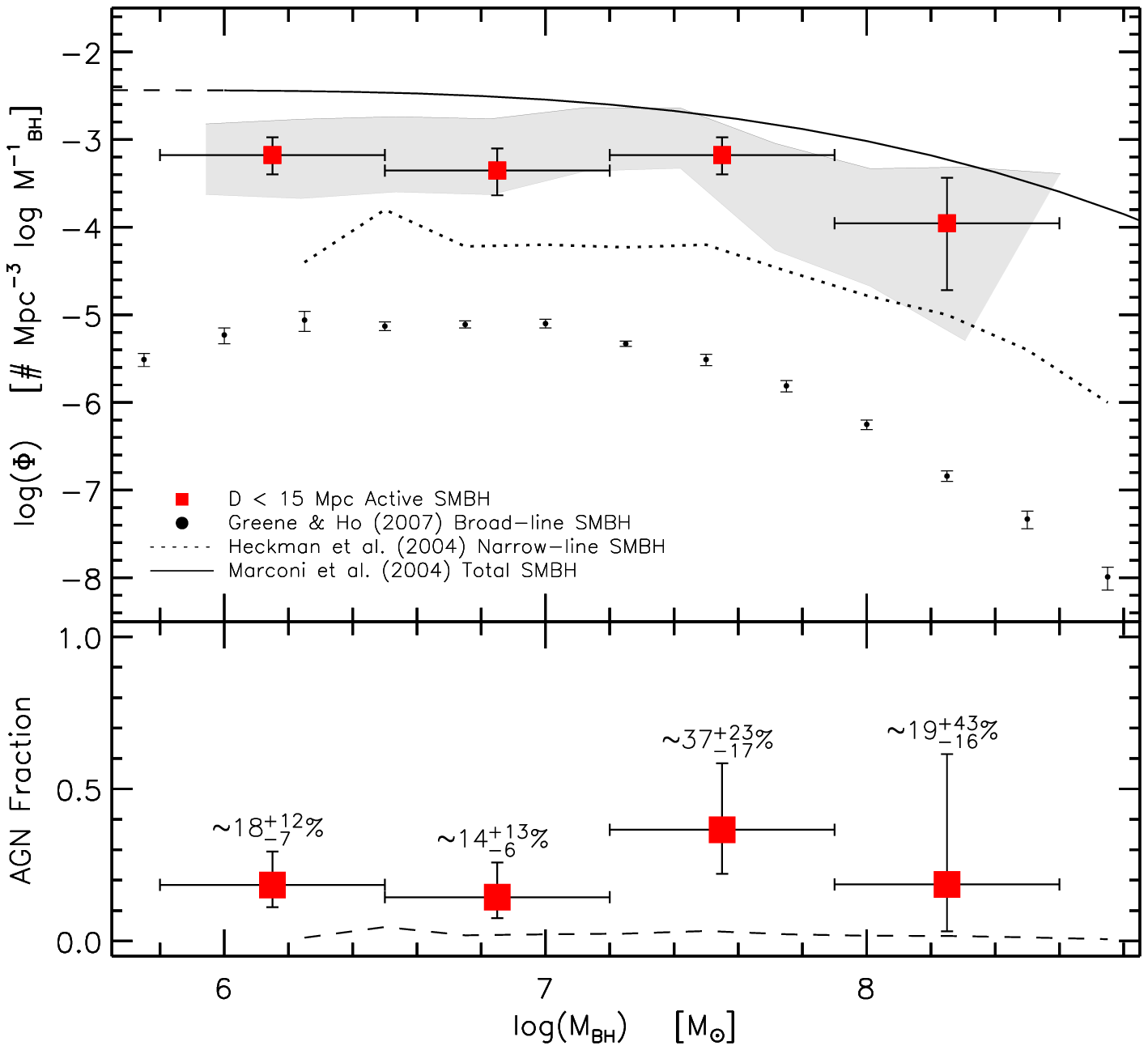}
\caption{(Upper panel) Comparison of volume-weighted space densities
  of active SMBHs in the local Universe, $\Phi$ in units of number
  ${\rm Mpc}^{-3} {\rm \ log \ M}^{-1}_{{\rm BH}}$. Mid-IR active SMBH
  function (filled squares; Goulding \& Alexander 2009) is compared to
  the optically identified NL AGN function (dotted curve) of H04, the
  BL AGN function (filled circle) of Greene \& Ho (2007), and total
  SMBH function (active+inactive galaxies; solid curve; Marconi
  et~al. 2004). Sample selection bias is analysed using a robust
  Monte-Carlo simulation (shaded region; see Appendix). (Lower panel)
  Ratio of mid-IR active SMBHs to the total local SMBH mass
  function. The total SMBH mass function is extrapolated by 0.3 dex to
  $\Mbh < 10^6 \Msun$. For comparison the volume-weighted AGN fraction
  of H04 is also shown (dashed line). We estimate a mean
  volume-weighted local AGN fraction of $\approx 25^{+29}_{-14}$
  percent over the range $\Mbh \approx (0.5$--$500) \times 10^6
  \Msun$.}
\label{fig_3}
\end{center}
\end{figure*}

Using the relative mass accretion rates estimated for our sample
(Fig.~\ref{fig_2}), we can infer the volume-averaged growth time of
SMBHs in the local Universe. Assuming a mean Kerr spin parameter ($a$)
for our sample of $a \approx 0.67$ (e.g.,
\citealt{treister06,hopkins07}), i.e., an accretion efficiency
($\epsilon$) of $\approx 0.1$, the characteristic mass doubling time
($t_{2M}$) of a SMBH accreting matter at the Eddington limit is
$t_{2M} \approx 30$~Myrs (Rees 1984). Under the further assumption
that $a$, and hence $\epsilon$, does not vary significantly for
changes in $\Mbh$ \citep{king08}, we assess the present-day growth
rate of SMBHs.\footnote{We note that the spin variation and spin
  directionality of SMBHs in AGNs is currently an ongoing area of
  research, and a consensus between groups has yet to be reached for
  an average value of the Kerr spin parameter; for example see
  \citet{brenneman06}, \citet{king08} and \citet{fabian09}.} Following
H04, we calculate and extend to lower masses ($\Mbh < 3 \times 10^6
\Msun$) the integrated growth of SMBHs. Growth time errors are
calculated from the log-normal standard deviations of the sample. We
note here that we also include the optically unidentified AGNs which
would not be detected in the SDSS.

In Fig.~\ref{fig_4}, we find that the mean growth time for low-mass
SMBHs ($\Mbh \approx 10^6 \Msun$) is $\approx 6^{+6}_{-3}$~Gyrs, which
is consistent with these AGNs growing on time-scales similar to that
of the age of the Universe. Our results are found to be broadly
consistent with a simple extrapolation of the growth times calculated
by H04 to $\Mbh \approx 10^6 \Msun$ (dashed-line in
Fig.~\ref{fig_4}). Thus, the AGNs hosting SMBHs in the mass range
$\Mbh \approx 10^6$--$10^7 \Msun$, which are dominated by optically
unidentified AGNs (see Fig.~\ref{fig_2}), are acquiring a significant
proportion of their mass in the present-day, and are amongst the most
rapidly growing SMBHs in the local Universe. Furthermore, we find our
derived growth times of SMBHs with $\Mbh \goa 3 \times 10^6 \Msun$ are
in good agreement with those presented in H04, with mean growth times
of $t_{2M} \approx 47^{+29}_{-18}$ and $\approx 198^{+195}_{-90}$~Gyrs
for AGNs with $\Mbh \approx (0.3$--3) and (3--$30) \times 10^7 \Msun$,
respectively.\footnote{We note that our sample contains only one
  galaxy with $\Mbh \goa 10^8 \Msun$ (NGC~5128) and thus may not be
  representative for high $\Mbh$ systems.}

\subsection{Space-Density of AGNs in the local Universe}

An accurate active SMBH mass function, especially for lower mass SMBHs
($\Mbh \approx 10^6 \Msun$), is crucial for extending our
understanding of the role played by accretion in the growth of SMBHs
across cosmic time. In this section we calculate the space density of
active SMBHs for our sample and compare it to complimentary optical
studies of local narrow-line (NL; H04) and broad-line (BL; Greene \&
Ho 2007) AGNs, and the total mass function of local SMBHs by Marconi
et~al. (2004).

Following Greene \& Ho (2007), in the top panel of Fig.~\ref{fig_3} we
plot the volume-weighted space density, $\Phi$ against $\Mbh$ in mass
bins of 0.5 dex. The volume, $V$, encompassed by the GA09 sample to
$D<15$~Mpc is $V \approx 1.3 \times 10^4$~Mpc$^3$.\footnote{The RBGS
  includes all IR-bright galaxies detected by {\it IRAS} with $f_{60
    \mu m}>5.24$ Jy at $|b|> 5 \degr$.} As we note in \S2, given the
luminosity limit imposed in our volume-limited survey ($L_{\rm IR}
\goa 3 \times 10^9 \Lsun$), we do not include the dwarf like systems
which are likely to host the very smallest SMBHs ($\Mbh < 10^5
\Msun$). With an adjustment for the distance model adopted in this
work, an examination of the Palomar survey \citep{ho97a} shows there
are possibly three optical Seyferts (NGC~185; NGC~1058; NGC~4395) with
$\Mbh < 10^5 \Msun$ which are not included in our sample due to our
lower luminosity limit. However, we also note that NGC~185 has since
been re-classified as an HII galaxy \citep{ho01b}.

 The derived volume-weighted space density for our active SMBHs
 (filled squares), which is dominated by NL-AGNs, is found to be
 significantly greater (a factor of $\approx 100$) than the SMBH
 density of BL-AGNs (filled circles) presented in Greene \& Ho (2007)
 in the mass region $\Mbh \approx (0.9$--$90) \times 10^{6}
 \Msun$. The significant increase in active SMBH density when compared
 to the BL-AGN density of Greene \& Ho (2007) is partially to be
 expected due to the greater relative sensitivity of our {\it
   Spitzer}-IRS observations coupled with the greater abundance of
 observed Seyfert 2 to Seyfert 1 galaxies identified in the local
 Universe. However, this still may not be a good indicator of the
 intrinsic Seyfert~1:Seyfert~2 ratio. \citet{tommasin10} find in a
 large sample of 81 Seyfert galaxies that only six sources do not
 contain significant \nev emission in their mid-IR spectroscopy, the
 majority of which are Seyfert~1s. It is likely that the
 identification of low equivalent-width emission lines (such as [NeV]
 or [OIV]) in BL-AGNs is further complicated by a strong IR continuum
 emission which dominates the mid-IR regime. Hence, it is possible
 that by requiring the detection of \nev to infer AGN status, we may
 be rejecting broad-line objects, and thus finding a lower
 Seyfert~1:Seyfert~2 ratio than is representative in the local
 Universe.

In comparison to the active SMBH mass function containing the
optically identified NL-AGNs of H04 (dotted line), we also find a
significantly larger space-density of SMBHs. We find that the
space-density of active SMBHs identified in the mid-IR is roughly
constant in the mass region $\Mbh \approx (0.9$--$90) \times 10^{6}
\Msun$ with a value of $\Phi \approx 6.3 \times 10^{-4} {\rm
  \ Mpc}^{-3} {\rm \ log M}^{-1}_{\rm BH}$. This space-density of AGNs
is a factor of $\approx 10$ greater than that estimated by H04 over
the same $\Mbh$ range. Since we find only two of the 17 AGNs in our
sample are sufficiently luminous/unobscured to be detected in the SDSS
survey, we determine that this is consistent with our results. We
further suggest that the space density derived here may still be a
lower-limit for the number of NL-AGNs in the local Universe. A further
examination of the (distance-model adjusted) Palomar survey suggests
that at least four further NL-AGNs are not included in our
volume-limited survey. Of these, two (NGC~3486; NGC~4565) lack
high-resolution {\it Spitzer}-IRS spectroscopy of the central region
(as noted in Table 2 of GA09), one (NGC~3031; $\Lir \approx 2.8 \times
10^9 \Lsun$) lies fractionally below our luminosity limit for this
survey, and NGC~4258 is not included in the RBGS due to its extremely
large angular size.

Whilst the main focus of this paper is to compare SMBH statistics
derived from mid-IR and optical detection techniques, it is prudent to
note that the majority of AGN space densities calculated at higher
redshifts are typically derived from sources detected in wide-field
X-ray surveys. Hence, we now establish whether our mid-IR active space
density may be missing a significant fraction of X-ray detected
AGNs. Recently, a comparison between high-resolution {\it Spitzer}-IRS
spectroscopy and X-ray detected AGNs, was made by \citet{dudik09} for
a large sample of optically classified LINERS. Dudik et~al. (2009)
reported inconsistencies between [NeV] non-detections and the presence
of hard X-ray nuclear emission in a subset of their sample. However,
they conclude that the limited sensitivity of their mid-IR
observations may be driving their observed result. Using the relations
between high-ionisation mid-IR emission and hard X-ray luminosity
(M08; GA09; Goulding et al. in preparation), we suggest that a
detected [NeV] luminosity of $L_{\rm [NeV]} \approx 10^{38} \ergps$ in
an AGN (i.e., the limiting luminosity in Dudik et~al. 2009) would be
equivalent to a hard X-ray luminosity of $L_{\rm X,2-10keV} \approx 5
\times 10^{40} \ergps$; indeed, almost all of the X-ray detected AGNs
which lack significant [NeV] emission in Dudik et~al. (2009) are below
this threshold. We thus conclude that with sufficiently sensitive
high-resolution mid-IR spectroscopy, there is currently no conclusive
evidence to suggest that X-ray detected Type-2 AGNs lack significant
[NeV] emission in the mid-IR. Hence, with the exception of some
Seyfert 1 galaxies (as discussed previously), it is unlikely that our
derived mid-IR space density lacks significant numbers of X-ray
detected AGNs with [NeV] emission below our sensitivity limit.

Given our comparatively small volume to that considered by using the
SDSS (e.g., Greene \& Ho 2007; H04), we further validate our derived
space density of active SMBHs by robustly testing our results to find
if: 1) our sample is over-dense, and thus strongly subject to cosmic
variance; or 2) given the modest errors associated with our $\Mbh$
estimations, the derived space-density is strongly subject to
scattering of objects in our defined binning structure. We discuss
these analyses in the Appendix. Briefly, we find that our sample is
broadly representative of galaxies to $z \sim 0.3$, and further show
that even in our most pessimistic case, we find an increase in our
derived space-density of at least a factor of $\approx 2$ (maximum
increase by a factor of $\approx 11$) at $\Mbh \approx 3 \times 10^6
\Msun$ over the optical NL space-density of H04.

\subsection{The volume-weighted local AGN fraction}

The ratio of the space densities of the active SMBH to total SMBH mass
function (i.e., the volume-weighted local active SMBH fraction) is
shown in the lower panel of Fig. \ref{fig_3}. We calculate an overall
active SMBH fraction of $\approx 25^{+29}_{-14}$ percent for SMBHs of
$\Mbh \approx (0.5$--$500) \times 10^6 \Msun$ down to our [NeV]
completeness limit ($L_{\rm [NeV]} \ga 10^{38} \ergps$). We find that
this fraction is consistent with being constant throughout this $\Mbh$
range. However, given our detection sensitivity limit, we are unable
to probe lower Eddington ratios for AGNs hosting smaller
SMBHs. Instead, we consider the effect of the AGN fraction for a fixed
value of Eddington ratio (e.g., $\eta > 10^{-3}$; i.e., thin-disc
accretion systems) and find tentative evidence that the AGN fraction
($\approx 16^{+9}_{-6}$ and $\approx 8^{+10}_{-5}$ percent) may increase
with decreasing SMBH mass ($\Mbh \approx 10^6$--$10^7 \Msun$ and $\Mbh
\approx 10^7$--$10^8 \Msun$ bins, respectively).

For the lowest-mass SMBHs ($\Mbh \approx (5$--$30) \times 10^5
\Msun$), we estimate an overall non-negligible volume-weighted AGN
fraction of $18^{+12}_{-7}$ percent, potentially showing that a
considerable proportion of small bulge (and pseudo-bulge) galaxies
(i.e., late-type spiral galaxies; Sc--Sd) host AGN activity. It has
been previously suggested by H04 and Greene \& Ho (2007) that the AGN
fraction may peak at $\Mbh \approx (0.7$--$2) \times 10^7
\Msun$. However, with the inclusion of the additional low-mass
optically unidentified AGNs (see Fig.~\ref{fig_2}), we find that the
AGN fractions are consistent with remaining constant or even
increasing for $\Mbh < 10^7 \Msun$. 

As noted in \S5.3, the space-density of AGNs, and hence the local AGN
fraction derived in this work, may only be a lower-limit given the
nature of our volume-limited survey which by definition does not
include IR-faint systems. To improve upon these current source
statistics, a larger sample of late-type spiral galaxies would be
required to investigate our findings further. With the greater
sensitivity and resolving power proposed for the next generation of
space-based mid-IR spectrographs, for example the {\it Space Infrared
  Telescope for Cosmology and Astrophysics (SPICA)}\footnote{See
  http://www.ir.isas.jaxa.jp/SPICA/} and the mid-IR instrument (MIRI)
on-board the {\it James Webb Space Telescope (JWST)}\footnote{See
  http://www.roe.ac.uk/uktac/consortium/miri/}, surveys such as these
can be continued and extended to study more distant (i.e., greater
volumes) and heavily obscured AGNs.

\section[]{Conclusions}

We have presented the mean growth times and volume-weighted space
density of active SMBHs in the local Universe. Our sample of 17 AGNs
was derived from a sensitive volume-limited mid-IR spectral survey of
all IR bright galaxies to $D < 15$ Mpc carried out using the NASA
\spitzer \ Space Telescope (see GA09 for further details on the sample
selection). The most accurate SMBH masses available for the objects
are compiled from a variety of sources. For the three AGNs without
published $\Mbh$ estimates, we use a bulge/disc decomposition method
to determine the bulge luminosity and hence a SMBH mass (see section
3). Our main findings are the following:

\begin{enumerate}
\renewcommand{\theenumi}{(\arabic{enumi})}

\item {Using combined mid-IR emission line and high-quality hard X-ray
  constraints, we have derived accurate measurements of the intrinsic
  luminosities of our sample of AGNs (see section 4). In conjunction
  with the well-established SMBH measurements from previous studies
  and our own estimates from our aforementioned bulge/disc
  decomposition method, we have assessed the relative mass accretion
  rates of our sample. Due to our high sensitivity and the ability to
  probe low SMBH masses, we find that significant mass accretion
  ($\eta > 10^{-3}$) occurs onto SMBHs with $\Mbh \approx 10^6 \Msun$,
  the majority of which would not be detected in even the most
  sensitive optical surveys. See section~5.1.}
\item {Using our derived relative mass accretion rates for the sample
  and assuming a typical accretion efficiency of $\epsilon \approx
  0.1$, we assessed the characteristic mean mass doubling times
  ($t_{2M}$) for AGNs in the very nearby Universe. For AGNs hosting
  SMBHs with $\Mbh \approx (0.5$--$50) \times 10^7 \Msun$ we find
  consistent growth times ($t_{2M} \approx 47$--198~Gyrs) with those
  of the NL-AGNs identified in the SDSS (H04). However, we also find
  that SMBHs with $\Mbh < 5 \times 10^6 \Msun$ (i.e., below the
  spectral resolution limit of the SDSS), are amongst the most rapidly
  growing SMBHs in the local Universe, with present-day growth times
  consistent with (and possibly less than) the current age of the
  Universe ($t_{2M} \approx 6^{+6}_{-3}$~Gyrs). See section~5.2.}
\item {To assess the incidence of this population of low mass, rapidly
  growing SMBHs, we constructed a local space density function of
  active SMBHs. We find that active SMBHs may be at least a factor
  $\approx 2$ more common than previously identified in NL-AGN surveys
  using SDSS data. Furthermore, we estimate a non-negligible space
  density for low mass SMBHs ($\Mbh \approx 10^6 \Msun$) of $\Phi
  \approx 6 \times 10^{-4} {\rm Mpc}^{-3} {\rm \ log \ M}^{-1}_{{\rm
      BH}}$ which is consistent with the space density of more massive
  active SMBHs ($\Mbh \approx 10^7 \Msun$; i.e., those previously
  determined to be the most rapidly accreting population of
  SMBHs). See section~5.3}
\item {Using a local total SMBH mass function (Marconi et~al. 2004),
  we estimate a mean volume-weighted local AGN fraction of $\approx
  23^{+22}_{-11}$ percent, which remains relatively constant in the
  mass range $\Mbh \approx (1$--$10) \times 10^6 \Msun$. However, when
  only considering the SMBHs with $\eta > 10^{-3}$ (i.e., radiatively
  efficient accretion systems), we find tentative evidence for an
  increasing AGN fraction ($\approx 16^{+9}_{-6}$ and $\approx
  8^{+10}_{-5}$ percent) with decreasing SMBH mass (for $\Mbh \approx
  10^6$--$10^7 \Msun$ and $\Mbh \approx 10^7$--$10^8 \Msun$,
  respectively). See section~5.4}
\end{enumerate}


\section*{Acknowledgments}

We thank the referee for a considered report which significantly
improved the paper. We also thank the Science \& Technologies
Facilities Council (ADG; BDL), the Royal Society (DMA), the Leverhulme
Trust (DMA; JRM) and the Einstein Postdoctoral Fellowship (grant
number PF9-00064; BDL) for funding. We would like to thank A. Edge,
J. Geach, J. Greene, J. Lucey, M. Norris, I. Smail and V. Wild for
useful conversations. The authors acknowledge the use of the {\small
  GALFIT} program of Peng et al. This research has made use of the
NASA/IPAC Extra-galactic Database (NED) and the NASA \spitzer \ Space
Telescope which are operated by the Jet Propulsion Laboratory,
California Institute of Technology, under contract with the National
Aeronautics and Space Administration. This publication also makes use
of data from the 2MASS Extended Mission Ancillary Products, a project
collaboration between The University of Massachusetts and the Infrared
Processing and Analysis Center (JPL/ Caltech).


\bibliography{bibtex1}

\begin{thebibliography}{}

\bibitem[\protect\citeauthoryear{{Allen}, {Groves}, {Dopita}, {Sutherland} \&
  {Kewley}}{{Allen} et~al.}{2008}]{allen08}
{Allen} M.~G.,  {Groves} B.~A.,  {Dopita} M.~A.,  {Sutherland} R.~S.,
  {Kewley} L.~J.,  2008, \apjs, 178, 20

\bibitem[\protect\citeauthoryear{{Alonso-Herrero}, {P{\'e}rez-Gonz{\'a}lez},
  {Rieke}, {Alexander}, {Rigby}, {Papovich}, {Donley} \&
  {Rigopoulou}}{{Alonso-Herrero} et~al.}{2008}]{alonso08}
{Alonso-Herrero} A.,  {P{\'e}rez-Gonz{\'a}lez} P.~G.,  {Rieke} G.~H.,
  {Alexander} D.~M.,  {Rigby} J.~R.,  {Papovich} C.,  {Donley} J.~L.,
  {Rigopoulou} D.,  2008, \apj, 677, 127

\bibitem[\protect\citeauthoryear{{Antonucci}}{{Antonucci}}{1993}]{antonucci93}
{Antonucci} R.,  1993, \araa, 31, 473

\bibitem[\protect\citeauthoryear{{Armus}, {Bernard-Salas}, {Spoon}, {Marshall},
  {Charmandaris} \& {et al.}}{{Armus} et~al.}{2006}]{armus06}
{Armus} L.,  {Bernard-Salas} J.,  {Spoon} H.~W.~W.,  {Marshall} J.~A.,
  {Charmandaris} V.,    {et al.} 2006, \apj, 640, 204

\bibitem[\protect\citeauthoryear{{Awaki} \& {et~al.}}{{Awaki} et~al.}{2008}]{awaki08}
{Awaki} H.,  {et~al.} 2008, \pasj, 60, 293

\bibitem[\protect\citeauthoryear{{Baldwin}, {Phillips} \&
  {Terlevich}}{{Baldwin} et~al.}{1981}]{bpt}
{Baldwin} J.~A.,  {Phillips} M.~M.,    {Terlevich} R.,  1981, \pasp, 93, 5

\bibitem[\protect\citeauthoryear{{Barth}, {Greene} \& {Ho}}{{Barth}
  et~al.}{2005}]{barth05}
{Barth} A.~J.,  {Greene} J.~E.,    {Ho} L.~C.,  2005, \apjl, 619, L151

\bibitem[\protect\citeauthoryear{{Barth}, {Ho} \& {Sargent}}{{Barth}
  et~al.}{2002}]{barth02}
{Barth} A.~J.,  {Ho} L.~C.,    {Sargent} W.~L.~W.,  2002, \aj, 124, 2607

\bibitem[\protect\citeauthoryear{{Barth}, {Strigari}, {Bentz}, {Greene} \&
  {Ho}}{{Barth} et~al.}{2009}]{barth09}
{Barth} A.~J.,  {Strigari} L.~E.,  {Bentz} M.~C.,  {Greene} J.~E.,    {Ho}
  L.~C.,  2009, \apj, 690, 1031

\bibitem[\protect\citeauthoryear{{Beckmann}, {Barthelmy}, {Courvoisier},
  {Gehrels}, {Soldi}, {Tueller} \& {Wendt}}{{Beckmann}
  et~al.}{2007}]{beckmann07}
{Beckmann} V.,  {Barthelmy} S.~D.,  {Courvoisier} T.~J.-L.,  {Gehrels} N.,
  {Soldi} S.,  {Tueller} J.,    {Wendt} G.,  2007, \aap, 475, 827

\bibitem[\protect\citeauthoryear{{Benson}, {D{\v z}anovi{\'c}}, {Frenk} \&
  {Sharples}}{{Benson} et~al.}{2007}]{benson07}
{Benson} A.~J.,  {D{\v z}anovi{\'c}} D.,  {Frenk} C.~S.,    {Sharples} R.,
  2007, \mnras, 379, 841

\bibitem[\protect\citeauthoryear{{Bentz}, {Peterson}, {Netzer}, {Pogge} \&
  {Vestergaard}}{{Bentz} et~al.}{2009}]{bentz09b}
{Bentz} M.~C.,  {Peterson} B.~M.,  {Netzer} H.,  {Pogge} R.~W.,
  {Vestergaard} M.,  2009, \apj, 697, 160

\bibitem[\protect\citeauthoryear{{Bentz}, {Peterson}, {Pogge} \&
  {Vestergaard}}{{Bentz} et~al.}{2009}]{bentz09a}
{Bentz} M.~C.,  {Peterson} B.~M.,  {Pogge} R.~W.,    {Vestergaard} M.,  2009,
  \apjl, 694, L166

\bibitem[\protect\citeauthoryear{{Bird}, {et~al.} \& {}}{{Bird}
  et~al.}{2007}]{bird07}
{Bird} A.~J.,  {et~al.}   {} 2007, \apjs, 170, 175

\bibitem[\protect\citeauthoryear{{Blandford} \& {McKee}}{{Blandford} \&
  {McKee}}{1982}]{blandford82}
{Blandford} R.~D.,  {McKee} C.~F.,  1982, \apj, 255, 419

\bibitem[\protect\citeauthoryear{{Brenneman} \& {Reynolds}}{{Brenneman} \&
  {Reynolds}}{2006}]{brenneman06}
{Brenneman} L.~W.,  {Reynolds} C.~S.,  2006, \apj, 652, 1028

\bibitem[\protect\citeauthoryear{{Buchanan}, {Gallimore}, {O'Dea}, {Baum},
  {Axon}, {Robinson}, {Elitzur} \& {Elvis}}{{Buchanan}
  et~al.}{2006}]{buchanan06}
{Buchanan} C.~L.,  {Gallimore} J.~F.,  {O'Dea} C.~P.,  {Baum} S.~A.,  {Axon}
  D.~J.,  {Robinson} A.,  {Elitzur} M.,    {Elvis} M.,  2006, \aj, 132, 401

\bibitem[\protect\citeauthoryear{{Cappi}, {et~al.} \& {}}{{Cappi}
  et~al.}{2006}]{cappi06}
{Cappi} M.,  {et~al.}   {} 2006, \aap, 446, 459

\bibitem[\protect\citeauthoryear{{Cowie}, {Barger}, {Bautz}, {Brandt} \&
  {Garmire}}{{Cowie} et~al.}{2003}]{cowie03}
{Cowie} L.~L.,  {Barger} A.~J.,  {Bautz} M.~W.,  {Brandt} W.~N.,    {Garmire}
  G.~P.,  2003, \apjl, 584, L57

\bibitem[\protect\citeauthoryear{{Dadina}}{{Dadina}}{2007}]{dadina07}
{Dadina} M.,  2007, \aap, 461, 1209

\bibitem[\protect\citeauthoryear{{Dale}, {Smith} \& {et~al.}}{{Dale}
  et~al.}{2009}]{dale09}
{Dale} D.~A.,  {Smith} J.~D.~T.,    {et~al.} 2009, \apj, 693, 1821

\bibitem[\protect\citeauthoryear{{Dasyra}, {Ho}, {Armus}, {Ogle}, {Helou},
  {Peterson}, {Lutz}, {Netzer} \& {Sturm}}{{Dasyra} et~al.}{2008}]{dasyra08}
{Dasyra} K.~M.,  {Ho} L.~C.,  {Armus} L.,  {Ogle} P.,  {Helou} G.,  {Peterson}
  B.~M.,  {Lutz} D.,  {Netzer} H.,    {Sturm} E.,  2008, \apjl, 674, L9

\bibitem[\protect\citeauthoryear{{de Vaucouleurs}, {de Vaucouleurs}, {Corwin}
  Jr., {Buta}, {Paturel} \& {Fouque}}{{de Vaucouleurs} et~al.}{1991}]{rc3}
{de Vaucouleurs} G.,  {de Vaucouleurs} A.,  {Corwin} Jr. H.~G.,  {Buta} R.~J.,
  {Paturel} G.,    {Fouque} P.,  1991, {Third Reference Catalogue of Bright
  Galaxies}.
Volume 1-3, XII, 2069 pp.~7 figs..~ Springer-Verlag Berlin Heidelberg New York

\bibitem[\protect\citeauthoryear{{Deo}, {Crenshaw}, {Kraemer}, {Dietrich},
  {Elitzur}, {Teplitz} \& {Turner}}{{Deo} et~al.}{2007}]{deo07}
{Deo} R.~P.,  {Crenshaw} D.~M.,  {Kraemer} S.~B.,  {Dietrich} M.,  {Elitzur}
  M.,  {Teplitz} H.,    {Turner} T.~J.,  2007, \apj, 671, 124

\bibitem[\protect\citeauthoryear{{Done}, {Madejski} \& {Smith}}{{Done}
  et~al.}{1996}]{done96}
{Done} C.,  {Madejski} G.~M.,    {Smith} D.~A.,  1996, \apjl, 463, L63+

\bibitem[\protect\citeauthoryear{{Dudik}, {Satyapal} \& {Marcu}}{{Dudik}
  et~al.}{2009}]{dudik09}
{Dudik} R.~P.,  {Satyapal} S.,    {Marcu} D.,  2009, \apj, 691, 1501

\bibitem[\protect\citeauthoryear{{Elvis}, {Risaliti}, {Nicastro}, {Miller},
  {Fiore} \& {Puccetti}}{{Elvis} et~al.}{2004}]{elvis04}
{Elvis} M.,  {Risaliti} G.,  {Nicastro} F.,  {Miller} J.~M.,  {Fiore} F.,
  {Puccetti} S.,  2004, \apjl, 615, L25

\bibitem[\protect\citeauthoryear{{Fabian}, {Zoghbi}, {Ross}, {Uttley}, {Gallo}
  \& {et~al.}}{{Fabian} et~al.}{2009}]{fabian09}
{Fabian} A.~C.,  {Zoghbi} A.,  {Ross} R.~R.,  {Uttley} P.,  {Gallo} L.~C.,
  {et~al.} 2009, \nat, 459, 540

\bibitem[\protect\citeauthoryear{{Ferrarese} \& {Merritt}}{{Ferrarese} \&
  {Merritt}}{2000}]{ferrarese00}
{Ferrarese} L.,  {Merritt} D.,  2000, \apjl, 539, L9

\bibitem[\protect\citeauthoryear{{Fukazawa}, {Iyomoto}, {Kubota}, {Matsumoto}
  \& {Makishima}}{{Fukazawa} et~al.}{2001}]{fukazawa01}
{Fukazawa} Y.,  {Iyomoto} N.,  {Kubota} A.,  {Matsumoto} Y.,    {Makishima} K.,
   2001, \aap, 374, 73

\bibitem[\protect\citeauthoryear{{Garcia-Rissmann}, {Vega}, {Asari}, {Cid
  Fernandes}, {Schmitt}, {Gonz{\'a}lez Delgado} \&
  {Storchi-Bergmann}}{{Garcia-Rissmann} et~al.}{2005}]{garcia05}
{Garcia-Rissmann} A.,  {Vega} L.~R.,  {Asari} N.~V.,  {Cid Fernandes} R.,
  {Schmitt} H.,  {Gonz{\'a}lez Delgado} R.~M.,    {Storchi-Bergmann} T.,  2005,
  \mnras, 359, 765

\bibitem[\protect\citeauthoryear{{Gebhardt}, {et~al.} \& {}}{{Gebhardt}
  et~al.}{2000}]{gebhardt00}
{Gebhardt} K.,  {et~al.}   {} 2000, \apjl, 539, L13

\bibitem[\protect\citeauthoryear{{Goulding} \& {Alexander}}{{Goulding} \&
  {Alexander}}{2009}]{GA09}
{Goulding} A.~D.,  {Alexander} D.~M.,  2009, \mnras, 398, 1165

\bibitem[\protect\citeauthoryear{{Greene} \& {Ho}}{{Greene} \&
  {Ho}}{2006}]{gre_Ho06}
{Greene} J.~E.,  {Ho} L.~C.,  2006, \apjl, 641, L21

\bibitem[\protect\citeauthoryear{{Greene} \& {Ho}}{{Greene} \&
  {Ho}}{2007}]{gre_ho07}
{Greene} J.~E.,  {Ho} L.~C.,  2007, \apj, 667, 131

\bibitem[\protect\citeauthoryear{{Greenhill}, {Gwinn}, {Antonucci} \&
  {Barvainis}}{{Greenhill} et~al.}{1996}]{greenhill96}
{Greenhill} L.~J.,  {Gwinn} C.~R.,  {Antonucci} R.,    {Barvainis} R.,  1996,
  \apjl, 472, L21+

\bibitem[\protect\citeauthoryear{{Greenhill}, {Moran} \&
  {Herrnstein}}{{Greenhill} et~al.}{1997}]{greenhill97}
{Greenhill} L.~J.,  {Moran} J.~M.,    {Herrnstein} J.~R.,  1997, \apjl, 481,
  L23+

\bibitem[\protect\citeauthoryear{{Guainazzi}, {Matt}, {Brandt}, {Antonelli},
  {Barr} \& {Bassani}}{{Guainazzi} et~al.}{2000}]{guainazzi00}
{Guainazzi} M.,  {Matt} G.,  {Brandt} W.~N.,  {Antonelli} L.~A.,  {Barr} P.,
  {Bassani} L.,  2000, \aap, 356, 463

\bibitem[\protect\citeauthoryear{{Guainazzi}, {Rodriguez-Pascual}, {Fabian},
  {Iwasawa} \& {Matt}}{{Guainazzi} et~al.}{2004}]{guainazzi04}
{Guainazzi} M.,  {Rodriguez-Pascual} P.,  {Fabian} A.~C.,  {Iwasawa} K.,
  {Matt} G.,  2004, \mnras, 355, 297

\bibitem[\protect\citeauthoryear{{Hao}, {Wu}, {Charmandaris}, {Spoon},
  {Bernard-Salas}, {Devost}, {Lebouteiller} \& {Houck}}{{Hao}
  et~al.}{2009}]{hao09}
{Hao} L.,  {Wu} Y.,  {Charmandaris} V.,  {Spoon} H.~W.~W.,  {Bernard-Salas} J.,
   {Devost} D.,  {Lebouteiller} V.,    {Houck} J.~R.,  2009, \apj, 704, 1159

\bibitem[\protect\citeauthoryear{{Hasinger}, {Miyaji} \& {Schmidt}}{{Hasinger}
  et~al.}{2005}]{hasinger05}
{Hasinger} G.,  {Miyaji} T.,    {Schmidt} M.,  2005, \aap, 441, 417

\bibitem[\protect\citeauthoryear{{H{\"a}ussler}, {McIntosh}, {Barden}, {Bell},
  {Rix} \& {et al}}{{H{\"a}ussler} et~al.}{2007}]{haussler07}
{H{\"a}ussler} B.,  {McIntosh} D.~H.,  {Barden} M.,  {Bell} E.~F.,  {Rix}
  H.-W.,    {et al} 2007, \apjs, 172, 615

\bibitem[\protect\citeauthoryear{{Heckman}, {Kauffmann}, {Brinchmann},
  {Charlot}, {Tremonti} \& {White}}{{Heckman} et~al.}{2004}]{heckman04}
{Heckman} T.~M.,  {Kauffmann} G.,  {Brinchmann} J.,  {Charlot} S.,  {Tremonti}
  C.,    {White} S.~D.~M.,  2004, \apj, 613, 109

\bibitem[\protect\citeauthoryear{{Ho}, {Filippenko} \& {Sargent}}{{Ho}
  et~al.}{1997a}]{ho97a}
{Ho} L.~C.,  {Filippenko} A.~V.,    {Sargent} W.~L.~W.,  1997a, \apjs, 112, 315

\bibitem[\protect\citeauthoryear{{Ho}, {Filippenko} \& {Sargent}}{{Ho}
  et~al.}{1997b}]{ho97b}
{Ho} L.~C.,  {Filippenko} A.~V.,    {Sargent} W.~L.~W.,  1997b, \apj, 487, 568

\bibitem[\protect\citeauthoryear{{Ho}, {Greene}, {Filippenko} \&
  {Sargent}}{{Ho} et~al.}{2009}]{ho09}
{Ho} L.~C.,  {Greene} J.~E.,  {Filippenko} A.~V.,    {Sargent} W.~L.~W.,  2009,
  \apjs, 183, 1

\bibitem[\protect\citeauthoryear{{Ho} \& {Ulvestad}}{{Ho} \&
  {Ulvestad}}{2001}]{ho01b}
{Ho} L.~C.,  {Ulvestad} J.~S.,  2001, \apjs, 133, 77

\bibitem[\protect\citeauthoryear{{Hopkins}, {Richards} \&
  {Hernquist}}{{Hopkins} et~al.}{2007}]{hopkins07}
{Hopkins} P.~F.,  {Richards} G.~T.,    {Hernquist} L.,  2007, \apj, 654, 731

\bibitem[\protect\citeauthoryear{{Israel}}{{Israel}}{1998}]{israel98}
{Israel} F.~P.,  1998, \aapr, 8, 237

\bibitem[\protect\citeauthoryear{{Itoh}, {Done}, {Makishima}, {Madejski} \&
  {et~al.}}{{Itoh} et~al.}{2008}]{itoh08}
{Itoh} T.,  {Done} C.,  {Makishima} K.,  {Madejski} G.,    {et~al.} 2008,
  \pasj, 60, 251

\bibitem[\protect\citeauthoryear{{Iwasawa}, {Koyama}, {Awaki}, {Kunieda},
  {Makishima}, {Tsuru}, {Ohashi} \& {Nakai}}{{Iwasawa}
  et~al.}{1993}]{iwasawa93}
{Iwasawa} K.,  {Koyama} K.,  {Awaki} H.,  {Kunieda} H.,  {Makishima} K.,
  {Tsuru} T.,  {Ohashi} T.,    {Nakai} N.,  1993, \apj, 409, 155

\bibitem[\protect\citeauthoryear{{Iyomoto}, {Fukazawa}, {Nakai} \&
  {Ishihara}}{{Iyomoto} et~al.}{2001}]{iyomoto01}
{Iyomoto} N.,  {Fukazawa} Y.,  {Nakai} N.,    {Ishihara} Y.,  2001, \apjl, 561,
  L69

\bibitem[\protect\citeauthoryear{{Jarrett}, {Chester}, {Cutri}, {Schneider} \&
  {Huchra}}{{Jarrett} et~al.}{2003}]{jarrett03}
{Jarrett} T.~H.,  {Chester} T.,  {Cutri} R.,  {Schneider} S.~E.,    {Huchra}
  J.~P.,  2003, \aj, 125, 525

\bibitem[\protect\citeauthoryear{{Kauffmann}, {Heckman}, {Tremonti} \&
  {et~al.}}{{Kauffmann} et~al.}{2003}]{kauff03b}
{Kauffmann} G.,  {Heckman} T.~M.,  {Tremonti} C.,    {et~al.} 2003, \mnras,
  346, 1055

\bibitem[\protect\citeauthoryear{{King}, {Pringle} \& {Hofmann}}{{King}
  et~al.}{2008}]{king08}
{King} A.~R.,  {Pringle} J.~E.,    {Hofmann} J.~A.,  2008, \mnras, 385, 1621

\bibitem[\protect\citeauthoryear{{Kormendy} \& {Kennicutt} Jr.}{{Kormendy} \&
  {Kennicutt}}{2004}]{kormendy04}
{Kormendy} J.,  {Kennicutt} Jr. R.~C.,  2004, \araa, 42, 603

\bibitem[\protect\citeauthoryear{{Kormendy} \& {Richstone}}{{Kormendy} \&
  {Richstone}}{1995}]{kormendy95}
{Kormendy} J.,  {Richstone} D.,  1995, \araa, 33, 581

\bibitem[\protect\citeauthoryear{{Madejski}, {{\.Z}ycki}, {Done}, {Valinia},
  {Blanco}, {Rothschild} \& {Turek}}{{Madejski} et~al.}{2000}]{madejski00}
{Madejski} G.,  {{\.Z}ycki} P.,  {Done} C.,  {Valinia} A.,  {Blanco} P.,
  {Rothschild} R.,    {Turek} B.,  2000, \apjl, 535, L87

\bibitem[\protect\citeauthoryear{{Magorrian}, {Tremaine}, {Richstone},
  {Bender}, {Bower}, {Dressler}, {Faber}, {Gebhardt}, {Green}, {Grillmair},
  {Kormendy} \& {Lauer}}{{Magorrian} et~al.}{1998}]{magorrian98}
{Magorrian} J.,  {Tremaine} S.,  {Richstone} D.,  {Bender} R.,  {Bower} G.,
  {Dressler} A.,  {Faber} S.~M.,  {Gebhardt} K.,  {Green} R.,  {Grillmair} C.,
  {Kormendy} J.,    {Lauer} T.,  1998, \aj, 115, 2285

\bibitem[\protect\citeauthoryear{{Maiolino}, {Salvati}, {Bassani}, {Dadina},
  {della Ceca}, {Matt}, {Risaliti} \& {Zamorani}}{{Maiolino}
  et~al.}{1998}]{maiolino98}
{Maiolino} R.,  {Salvati} M.,  {Bassani} L.,  {Dadina} M.,  {della Ceca} R.,
  {Matt} G.,  {Risaliti} G.,    {Zamorani} G.,  1998, \aap, 338, 781

\bibitem[\protect\citeauthoryear{{Marconi}, {Capetti}, {Axon}, {Koekemoer},
  {Macchetto} \& {Schreier}}{{Marconi} et~al.}{2001}]{marconi01}
{Marconi} A.,  {Capetti} A.,  {Axon} D.~J.,  {Koekemoer} A.,  {Macchetto} D.,
   {Schreier} E.~J.,  2001, \apj, 549, 915

\bibitem[\protect\citeauthoryear{{Marconi}, {Risaliti}, {Gilli}, {Hunt},
  {Maiolino} \& {Salvati}}{{Marconi} et~al.}{2004}]{marconi04}
{Marconi} A.,  {Risaliti} G.,  {Gilli} R.,  {Hunt} L.~K.,  {Maiolino} R.,
  {Salvati} M.,  2004, \mnras, 351, 169

\bibitem[\protect\citeauthoryear{{Matsumoto}, {Nava}, {Maddox}, {Leighly},
  {Grupe}, {Awaki} \& {Ueno}}{{Matsumoto} et~al.}{2004}]{matsumoto04}
{Matsumoto} C.,  {Nava} A.,  {Maddox} L.~A.,  {Leighly} K.~M.,  {Grupe} D.,
  {Awaki} H.,    {Ueno} S.,  2004, \apj, 617, 930

\bibitem[\protect\citeauthoryear{{Matt}, {et~al.}, {} \& {}}{{Matt}
  et~al.}{1997}]{matt97}
{Matt} G.,  {et~al.} {}   {} 1997, \aap, 325, L13

\bibitem[\protect\citeauthoryear{{McLure} \& {Dunlop}}{{McLure} \&
  {Dunlop}}{2004}]{mclure04}
{McLure} R.~J.,  {Dunlop} J.~S.,  2004, \mnras, 352, 1390

\bibitem[\protect\citeauthoryear{{Mel{\'e}ndez}, {Kraemer}, {Armentrout},
  {Deo}, {Crenshaw}, {Schmitt}, {Mushotzky}, {Tueller}, {Markwardt} \&
  {Winter}}{{Mel{\'e}ndez} et~al.}{2008}]{melendez08b}
{Mel{\'e}ndez} M.,  {Kraemer} S.~B.,  {Armentrout} B.~K.,  {Deo} R.~P.,
  {Crenshaw} D.~M.,  {Schmitt} H.~R.,  {Mushotzky} R.~F.,  {Tueller} J.,
  {Markwardt} C.~B.,    {Winter} L.,  2008, \apj, 682, 94

\bibitem[\protect\citeauthoryear{{Mould}, {Huchra} \& {et~al.}}{{Mould}
  et~al.}{2000}]{mould00}
{Mould} J.~R.,  {Huchra} J.~P.,    {et~al.} 2000, \apj, 529, 786

\bibitem[\protect\citeauthoryear{{Narayan} \& {Yi}}{{Narayan} \&
  {Yi}}{1994}]{adaf}
{Narayan} R.,  {Yi} I.,  1994, \apjl, 428, L13

\bibitem[\protect\citeauthoryear{{Nelson}, {Green}, {Bower}, {Gebhardt} \&
  {Weistrop}}{{Nelson} et~al.}{2004}]{nelson04}
{Nelson} C.~H.,  {Green} R.~F.,  {Bower} G.,  {Gebhardt} K.,    {Weistrop} D.,
  2004, \apj, 615, 652

\bibitem[\protect\citeauthoryear{{Onken}, {Ferrarese}, {Merritt}, {Peterson},
  {Pogge}, {Vestergaard} \& {Wandel}}{{Onken} et~al.}{2004}]{onken04}
{Onken} C.~A.,  {Ferrarese} L.,  {Merritt} D.,  {Peterson} B.~M.,  {Pogge}
  R.~W.,  {Vestergaard} M.,    {Wandel} A.,  2004, \apj, 615, 645

\bibitem[\protect\citeauthoryear{{Onken}, {Peterson}, {Dietrich}, {Robinson} \&
  {Salamanca}}{{Onken} et~al.}{2003}]{onken03}
{Onken} C.~A.,  {Peterson} B.~M.,  {Dietrich} M.,  {Robinson} A.,
  {Salamanca} I.~M.,  2003, \apj, 585, 121

\bibitem[\protect\citeauthoryear{{Peng}, {Ho}, {Impey} \& {Rix}}{{Peng}
  et~al.}{2002}]{galfit}
{Peng} C.~Y.,  {Ho} L.~C.,  {Impey} C.~D.,    {Rix} H.-W.,  2002, \aj, 124, 266

\bibitem[\protect\citeauthoryear{{Peterson} \& {Wandel}}{{Peterson} \&
  {Wandel}}{1999}]{peterson99}
{Peterson} B.~M.,  {Wandel} A.,  1999, \apjl, 521, L95

\bibitem[\protect\citeauthoryear{{Pounds}, {Reeves}, {King} \& {Page}}{{Pounds}
  et~al.}{2004}]{pounds04}
{Pounds} K.~A.,  {Reeves} J.~N.,  {King} A.~R.,    {Page} K.~L.,  2004, \mnras,
  350, 10

\bibitem[\protect\citeauthoryear{{Ratnam} \& {Salucci}}{{Ratnam} \&
  {Salucci}}{2000}]{ratnam00}
{Ratnam} C.,  {Salucci} P.,  2000, New Astronomy, 5, 427

\bibitem[\protect\citeauthoryear{{Rees}}{{Rees}}{1984}]{rees84}
{Rees} M.~J.,  1984, \araa, 22, 471

\bibitem[\protect\citeauthoryear{{Risaliti}, {Salvati}, {Elvis}, {Fabbiano},
  {Baldi}, {Bianchi}, {Braito}, {Guainazzi}, {Matt}, {Miniutti}, {Reeves},
  {Soria} \& {Zezas}}{{Risaliti} et~al.}{2009}]{risaliti09}
{Risaliti} G.,  {Salvati} M.,  {Elvis} M.,  {Fabbiano} G.,  {Baldi} A.,
  {Bianchi} S.,  {Braito} V.,  {Guainazzi} M.,  {Matt} G.,  {Miniutti} G.,
  {Reeves} J.,  {Soria} R.,    {Zezas} A.,  2009, \mnras, 393, L1

\bibitem[\protect\citeauthoryear{{Roberts}, {Schurch} \& {Warwick}}{{Roberts}
  et~al.}{2001}]{roberts01}
{Roberts} T.~P.,  {Schurch} N.~J.,    {Warwick} R.~S.,  2001, \mnras, 324, 737

\bibitem[\protect\citeauthoryear{{Sanders}, {Mazzarella}, {Kim}, {Surace} \&
  {Soifer}}{{Sanders} et~al.}{2003}]{RBGS}
{Sanders} D.~B.,  {Mazzarella} J.~M.,  {Kim} D.-C.,  {Surace} J.~A.,
  {Soifer} B.~T.,  2003, \aj, 126, 1607

\bibitem[\protect\citeauthoryear{{Satyapal}, {Vega}, {Dudik}, {Abel} \&
  {Heckman}}{{Satyapal} et~al.}{2008}]{sat08}
{Satyapal} S.,  {Vega} D.,  {Dudik} R.~P.,  {Abel} N.~P.,    {Heckman} T.,
  2008, \apj, 677, 926

\bibitem[\protect\citeauthoryear{{Satyapal}, {Vega}, {Heckman}, {O'Halloran} \&
  {Dudik}}{{Satyapal} et~al.}{2007}]{sat07}
{Satyapal} S.,  {Vega} D.,  {Heckman} T.,  {O'Halloran} B.,    {Dudik} R.,
  2007, \apjl, 663, L9

\bibitem[\protect\citeauthoryear{{Schaerer} \& {Stasi{\'n}ska}}{{Schaerer} \&
  {Stasi{\'n}ska}}{1999}]{schaerer99}
{Schaerer} D.,  {Stasi{\'n}ska} G.,  1999, \aap, 345, L17

\bibitem[\protect\citeauthoryear{{Shirai}, {Fukazawa}, {Sasada}, {Ohno},
  {Yonetoku}, {Yokota}, {Fujimoto}, {Murakami}, {Terashima}, {Awaki}, {Ikeda},
  {Ozawa} \& {Tsuru}}{{Shirai} et~al.}{2008}]{shirai08}
{Shirai} H.,  {Fukazawa} Y.,  {Sasada} M.,  {Ohno} M.,  {Yonetoku} D.,
  {Yokota} S.,  {Fujimoto} R.,  {Murakami} T.,  {Terashima} Y.,  {Awaki} H.,
  {Ikeda} S.,  {Ozawa} M.,    {Tsuru} T.~G.,  2008, \pasj, 60, 263

\bibitem[\protect\citeauthoryear{{Soltan}}{{Soltan}}{1982}]{soltan82}
{Soltan} A.,  1982, \mnras, 200, 115

\bibitem[\protect\citeauthoryear{{Tanaka}, {Inoue} \& {Holt}}{{Tanaka}
  et~al.}{1994}]{tanaka94}
{Tanaka} Y.,  {Inoue} H.,    {Holt} S.~S.,  1994, \pasj, 46, L37

\bibitem[\protect\citeauthoryear{{Tommasin}, {Spinoglio}, {Malkan} \&
  {Fazio}}{{Tommasin} et~al.}{2010}]{tommasin10}
{Tommasin} S.,  {Spinoglio} L.,  {Malkan} M.~A.,    {Fazio} G.,  2010, \apj,
  709, 1257

\bibitem[\protect\citeauthoryear{{Treister} \& {Urry}}{{Treister} \&
  {Urry}}{2006}]{treister06}
{Treister} E.,  {Urry} C.~M.,  2006, \apjl, 652, L79

\bibitem[\protect\citeauthoryear{{Tremaine}, {Gebhardt}, {Bender}, {Bower},
  {Dressler}, {Faber}, {Filippenko}, {Green}, {Grillmair}, {Ho}, {Kormendy},
  {Lauer}, {Magorrian}, {Pinkney} \& {Richstone}}{{Tremaine}
  et~al.}{2002}]{tremaine02}
{Tremaine} S.,  {Gebhardt} K.,  {Bender} R.,  {Bower} G.,  {Dressler} A.,
  {Faber} S.~M.,  {Filippenko} A.~V.,  {Green} R.,  {Grillmair} C.,  {Ho}
  L.~C.,  {Kormendy} J.,  {Lauer} T.~R.,  {Magorrian} J.,  {Pinkney} J.,
  {Richstone} D.,  2002, \apj, 574, 740

\bibitem[\protect\citeauthoryear{{Tueller}, {Mushotzky}, {Barthelmy},
  {Cannizzo}, {Gehrels}, {Markwardt}, {Skinner} \& {Winter}}{{Tueller}
  et~al.}{2008}]{tueller08}
{Tueller} J.,  {Mushotzky} R.~F.,  {Barthelmy} S.,  {Cannizzo} J.~K.,
  {Gehrels} N.,  {Markwardt} C.~B.,  {Skinner} G.~K.,    {Winter} L.~M.,  2008,
  \apj, 681, 113

\bibitem[\protect\citeauthoryear{{Ueda}, {Akiyama}, {Ohta} \& {Miyaji}}{{Ueda}
  et~al.}{2003}]{ueda03}
{Ueda} Y.,  {Akiyama} M.,  {Ohta} K.,    {Miyaji} T.,  2003, \apj, 598, 886

\bibitem[\protect\citeauthoryear{{Valdes}, {Gupta}, {Rose}, {Singh} \&
  {Bell}}{{Valdes} et~al.}{2004}]{valdes04}
{Valdes} F.,  {Gupta} R.,  {Rose} J.~A.,  {Singh} H.~P.,    {Bell} D.~J.,
  2004, \apjs, 152, 251

\bibitem[\protect\citeauthoryear{{Vasudevan} \& {Fabian}}{{Vasudevan} \&
  {Fabian}}{2009}]{vasudevan09b}
{Vasudevan} R.~V.,  {Fabian} A.~C.,  2009, \mnras, 392, 1124

\bibitem[\protect\citeauthoryear{{Vasudevan}, {Fabian}, {Gandhi}, {Winter} \&
  {Mushotzky}}{{Vasudevan} et~al.}{2010}]{vasudevan10}
{Vasudevan} R.~V.,  {Fabian} A.~C.,  {Gandhi} P.,  {Winter} L.~M.,
  {Mushotzky} R.~F.,  2010, \mnras, 402, 1081

\bibitem[\protect\citeauthoryear{{Vasudevan}, {Mushotzky}, {Winter} \&
  {Fabian}}{{Vasudevan} et~al.}{2009}]{vasudevan09a}
{Vasudevan} R.~V.,  {Mushotzky} R.~F.,  {Winter} L.~M.,    {Fabian} A.~C.,
  2009, \mnras, pp 1234--+

\bibitem[\protect\citeauthoryear{{Vignati} \& {et~al.}}{{Vignati} et~al.}{1999}]{vignati99}
{Vignati} P.,  {et~al.} 1999, \aap, 349, L57

\bibitem[\protect\citeauthoryear{{Wandel}}{{Wandel}}{1999}]{wandel99a}
{Wandel} A.,  1999, \apjl, 519, L39

\bibitem[\protect\citeauthoryear{{Wandel}, {Peterson} \& {Malkan}}{{Wandel}
  et~al.}{1999}]{wandel99b}
{Wandel} A.,  {Peterson} B.~M.,    {Malkan} M.~A.,  1999, \apj, 526, 579

\bibitem[\protect\citeauthoryear{{Weedman}, {Hao}, {Higdon}, {Devost}, {Wu},
  {Charmandaris}, {Brandl}, {Bass} \& {Houck}}{{Weedman}
  et~al.}{2005}]{weedman05}
{Weedman} D.~W.,  {Hao} L.,  {Higdon} S.~J.~U.,  {Devost} D.,  {Wu} Y.,
  {Charmandaris} V.,  {Brandl} B.,  {Bass} E.,    {Houck} J.~R.,  2005, \apj,
  633, 706

\bibitem[\protect\citeauthoryear{{Whittle}}{{Whittle}}{1992}]{whittle92}
{Whittle} M.,  1992, \apjs, 79, 49

\bibitem[\protect\citeauthoryear{{Winter}, {Mushotzky}, {Reynolds} \&
  {Tueller}}{{Winter} et~al.}{2009}]{winter09}
{Winter} L.~M.,  {Mushotzky} R.~F.,  {Reynolds} C.~S.,    {Tueller} J.,  2009,
  \apj, 690, 1322

\bibitem[\protect\citeauthoryear{{Yang}, {Wilson}, {Matt}, {Terashima} \&
  {Greenhill}}{{Yang} et~al.}{2009}]{yang09}
{Yang} Y.,  {Wilson} A.~S.,  {Matt} G.,  {Terashima} Y.,    {Greenhill} L.~J.,
  2009, \apj, 691, 131

\bibitem[\protect\citeauthoryear{{Yaqoob} \& {et~al.}}{{Yaqoob} et~al.}{2007}]{yaqoob07}
{Yaqoob} T.,  {et~al.} 2007, \pasj, 59, 283

\bibitem[\protect\citeauthoryear{{York}, {Adelman}, {Anderson} Jr., {Anderson}
  \& {et~al}}{{York} et~al.}{2000}]{sdss_tech}
{York} D.~G.,  {Adelman} J.,  {Anderson} Jr. J.~E.,  {Anderson} S.~F.,
  {et~al} 2000, \aj, 120, 1579

\end{thebibliography}

\bsp 

\newpage
\appendix
\section{Validation of derived space-density of active SMBHs}

As the space-density of active SMBHs derived in this work is
significantly larger than that found in previous studies, in this
section we attempt to validate our results by discussing possible
limitations and additional sources of error which may exist in these
analyses: 1) whilst the sensitivity of the data used in this survey is
high, the volume considered is relatively small compared to that of
the SDSS, thus our results may be subject to cosmic variance; 2) given
the modest errors associated with the $\Mbh$ estimates, the adopted
$\Mbh$ binning structure is likely to be subjective and thus
degenerate towards objects scattering between the defined bins. Under
these assumptions, in \S A1 we investigate whether the sample is
indeed representative of the local Universe and in \S A2 we discuss
the construction of a Monte Carlo simulation to assess the effect of
our adopted $\Mbh$ binning.

\subsection{Is our sample representative of the local Universe?}

Given the large incidence of AGNs within our sample it is possible
that the volume considered in our sample is over-dense compared to
other regions in the local Universe. The construction of the original
sample of 64 bolometrically luminous galaxies in GA09 was designed to
be complete down to the flux-limit of the Revised Bright Galaxy
Sample. This imposed a distance constraint of $D<15$ Mpc (see Fig. 1
of GA09), and hence did not include the Virgo cluster at $D \approx
16$ Mpc, and thus our sample does not incorporate known local
over-densities; however, this volume may not be representative of the
Universe at large.

To robustly test our considered volume ($V \approx 1.3 \times
10^4$~Mpc$^3$), we constructed a total SMBH space density function for
all galaxies to $D<15$ Mpc and compared this to the local total
(active+inactive) SMBH mass function of Marconi et~al. (2004) derived
from the luminosity function of local galaxies. Given the large
co-moving volume ($V_c \approx 1000 {\rm \ Gpc}^3$) considered in
Marconi et~al. (2004), their derived SMBH mass function is unlikely to
suffer from significant cosmic variance.

Our total SMBH mass function was formulated using all galaxies
identified in the NASA/IPAC Extra-galactic Database (NED) to
$D<15$~Mpc with a total {\it K}-band luminosity of $L_{\rm K,gal} \ga
1.5 \times 10^9 \Lsun$. The luminosity threshold is designed to
include all galaxies which could potentially host a SMBH with $\Mbh
\ga 10^6 \Msun$ using the $\Mbh$--$L_{\rm K,bul}$ relation. This
conservative lower limit assumes that $L_{\rm K,gal}= L_{\rm K,bul}$
(i.e., that all galaxies in the sample have an early-type galaxy
classification). In reality, the majority of the sources identified in
NED to $D<15$~Mpc are late-type galaxies (i.e., $L_{\rm K,gal} \gg
L_{\rm K,bul}$), and therefore the $\Mbh$ limit is likely to be $\Mbh
\ll 10^6 \Msun$. To $D<15$ Mpc, we identify 105 galaxies which
potentially host a SMBH with $\Mbh \goa 10^6 \Msun$. To estimate $\Mbh$
in each of these galaxies, we relate the associated Hubble-type from
the Third Reference Catalogue of Bright Galaxies \citep{rc3} to a mean
bulge/disc ratio (e.g., \citealt{benson07}) and establish an
individual bulge luminosity based on galaxy-type and $L_{\rm
  K,bul}$. We convert the estimated bulge luminosity to $\Mbh$ using
the $\Mbh$--$L_{\rm K,bul}$ relation and construct a total SMBH mass
function. Whilst this rather crude estimation carries large associated
errors, we still find very good agreement (a mean variance of 0.1 dex)
with the SMBH mass function of Marconi et al. (2004) throughout the
mass range $\Mbh \approx 10^6$--$10^9 \Msun$. Thus, to first-order,
our volume-limited sample does not appear to be over-dense and/or
subject to strong cosmic variance, and hence, is broadly
representative of the typical field-galaxy population in the local
Universe.

We further validate this conclusion by estimating the active SMBH mass
function for those AGNs which are sufficiently optically bright to be
included in H04. Under this assumption, the derived space densities of
active SMBHs for this sample and that of H04 should be comparable if
our sample is indeed representative of a field-galaxy population. We
find that using the H04 detection limit inferred from
Fig.~\ref{fig_2}, our new estimated space density is consistent with
H04 ($\Phi \approx 10^{-4}$) throughout the mass range considered here
($\Mbh \approx (0.05$--$30) \times 10^8 \Msun$), although there are
still considerable uncertainties given the small number statistics
inherent with our sample.

\subsection{Monte Carlo Analysis of $\Mbh$ binning}

We have established that our sample and the estimate of the SMBH mass
density (both from active and inactive galaxies) do not appear to be
subject to over-densities caused by cosmic variance. We therefore now
investigate the effect of small-number statistics on our results which
are inherent in relatively small samples such as this.

We show in \S3 that accurate estimates of $\Mbh$ for our sample of
AGNs from a homogeneous method is difficult to achieve. Thus, our
adopted values of $\Mbh$ are derived from a variety of
methodologies. In Fig.~\ref{fig_3}, the AGNs in the sample are placed
into equal bins of $\Mbh$ with width 0.5 dex. However, this binning
process does not allow for the error inherent to each individual
$\Mbh$ measurement. Hence, some objects may scatter out of one defined
bin and into another. The bin width we employ in our relatively modest
sample (64 objects) may therefore be subjective and requires
testing. Here we use a Monte Carlo analysis to assess the effect of
the scattering of AGNs into different $\Mbh$ bins on the derived space
density of active SMBHs in the local Universe~($\Phi$).

Our Monte Carlo analysis calculates $\Phi$ by selecting a random set
of SMBH masses from gaussian probability distributions constructed
using our adopted $\Mbh$ masses and their associated 1-$\sigma$ errors
given in column 12 of Table 1. The binning structure was designed to
incorporate at least one object in each bin from 250,000 realisations
of the simulation across the considered mass range of the sample
($\Mbh \approx (0.5$--$500) \times 10^6 \Msun$). We impose the upper
mass limit as we have only one high-mass AGN (NGC~5128; $\Mbh \approx
2.4 \times 10^8 \Msun$) in the sample. Thus, we determine that given
our sample distribution the maximum number of equal width bins to be
nine (i.e., equal bin sizes of 0.33 dex with $\goa 1$ object). Hence,
we use our simulation to conservatively assess the maximum error on
our calculated space density of active SMBHs.

The error shown in Fig.~\ref{fig_3} (shaded region) is the standard
deviation of the 250,000 simulations in our volume-limited sample
combined in quadrature with the Poisson error determined from the
number counts in our real sample. We found 250,000 realisations to be
sufficient, since at this level the maximum variation in the standard
deviation over multiple runs was less than $10^{-6}$.

We find that the mean spread in the derived value of $\Phi$ from our
volume-limited sample is $\approx 0.85$ dex, and as predicted appears
to be subject to some scattering of $\Mbh$. However, using our
conservative error analyses, we show that the space density of active
SMBHs found from our sample is consistently greater than that found
for optical narrow-line AGNs (H04) with $\Mbh < 10^8 \Msun$. Thus, to
first-order, we find a significant increase (of a minimum factor of
$\approx 2$ and a maximum of $\approx 80$) in the space density of
active SMBHs in the local Universe in the mass range $\Mbh \approx
(0.5$--$100) \times 10^6 \Msun$. For $\Mbh > 10^8 \Msun$ our errors
increase significantly due to very limited source statistics (see
\S4.1). We thus conclude that while this survey is subject to the
scattering of objects through the $\Mbh$ bins, even in our most
pessimistic case, we still find a significant increase in the space
density of active SMBHs in the local Universe compared to that found
in large-scale optical NL AGN studies.

\label{lastpage}

\end{document}